\def\etal{{\it et al.}}

\def\half{{\textstyle{1\over2}}}
\def\thalf{{\textstyle{3\over2}}}

\def\hlf{{{1\over2}}}
\def\thlf{{{3\over2}}}
\def\fhlf{{{5\over2}}}
\def\shlf{{{7\over2}}}
\def\>{\rangle}
\def\<{\langle}

\def\rmb#1{{\bf #1}}

\def\beq{\begin{equation}}
\def\eeq{\end{equation}}

\def\beqy{\begin{eqnarray}}
\def\eeqy{\end{eqnarray}}
\def\l{\lambda} 
\documentclass[aps,showpacs,floatfix]{revtex4}
\usepackage{graphicx,epsfig,rawfonts,pictex,color,multirow,longtable}
\newlength{\dinwidth}
\newlength{\dinmargin}
\begin{document}
\thispagestyle{empty}
\title{Heavy Baryons in a Quark Model\footnotetext{Notice: Authored by Jefferson Science Associates, LLC under U.S. DOE Contract No. DE-AC05-06OR23177. The U.S. Government retains a non-exclusive, paid-up, irrevocable, world-wide license to publish or reproduce this manuscript for U.S. Government purposes.}}
\author{W. Roberts$^{1}$ and Muslema Pervin$^{2}$}
\affiliation{$^1$ Department of Physics, Florida State University, Tallahassee, FL 32306\\
$^2$ Physics Division,~Argonne National Laboratory, Argonne, IL-60439}

\begin{abstract}
A quark model is applied to the spectrum of baryons containing heavy quarks. The model gives masses for the known heavy baryons that are in agreement with experiment, but for the doubly-charmed baryon $\Xi_{cc}$, the model prediction is too heavy. Mixing between the $\Xi_Q$ and $\Xi_Q^\prime$ states is examined and is found to be small for the lowest lying states. In contrast with this, mixing between the $\Xi_{bc}$ and $\Xi_{bc}^\prime$ states is found to be large, and the implication of this mixing for properties of these states is briefly discussed. We also examine heavy-quark spin-symmetry multiplets, and find that many states in the model can be placed in such multiplets. We compare our predictions with those of a number of other authors.
\end{abstract}
\maketitle
\begin{flushright}JLAB-THY-07-751\end{flushright}
\setcounter{page}{1}

%%%%%%%%%%%%%%%%%%%%%%%%%%%%%%%%%%%%%%%%%%%%%%%%%
\section{Introduction and Motivation}

\label{sec:intro}

Baryons containing heavy quarks have been the focus of much attention, particularly since the development of the heavy quark effective theory and its application to baryons containing a single heavy quark. One reason for this is that the heavy quark provides a `flavor tag' that may be used as a window into the heart and soul of confinement, or at least, a window that allows us to see somewhat further under the skin of nonperturbative QCD than do the light baryons. All of the states containing heavy quarks are expected to be somewhat narrow, for the most part, so that their detection and isolation is relatively easy, and in general does not rely on the extensive partial-wave-analysis machinery usually necessary for identifying light baryons (most states found to date have widths of a few MeV, with the largest reported width being a few tens of MeV, but with large uncertainties). Such analyses may be required for determining the quantum numbers of the states, but even there, the procedure may still be simpler than in the case of light baryons, as it is expected to be largely free of the various interferences that arise with nearby, broad and overlapping resonances.

In addition, the heavy quark symmetries provide a framework for understanding and predicting the spectrum of one flavor of heavy baryons, say those containing a $b$ quark, if the spectrum of baryons containing a $c$ quark has been obtained \cite{Isgur:1991wq}. Used judiciously, this heavy quark symmetry can provide some qualitative insight, and perhaps even quantitative, into the spectrum of light baryons, particularly the hyperons. 

Despite the wealth of information that they can provide, and many theoretical treatises, surprisingly little is known experimentally about the heavy baryons \cite{pdg}. This is largely because despite the comments above, they are difficult to produce. Unlike the heavy mesons, there are no resonant production mechanisms, so these baryons can only be obtained by continuum production, where cross sections are small, as products in the decays of heavy mesons, or at hadron colliders. Not surprisingly, the $B$ factories, and CLEO before that, have been the main source of these baryons, along with some recent contributions from the Fermilab Collaborations.

The known heavy baryons are shown in Table \ref{charm1}. None of the quantum numbers assigned in that table have been measured experimentally, but are based on quark model expectations. In a few cases, some guidance has been provided by a few of the decays of the baryons. For instance, the $\Lambda_c$ state at 2.880 GeV has been conjectured to have either $J^P=\frac{1}{2}^-$ \cite{Artuso:2000xy} or $J=\frac{5}{2}^\pm$ \cite{abe}. A very useful summary of the status of these baryons is given in \cite{chenga,cheng}. As can be seen from the table, there is much to be learned about the baryons with a single charm quark, while even less is known about the analogous baryons containing a single $b$ quark. To date there is only one candidate for a baryon containing more than one heavy quark. This is the $\Xi_{cc}$ at 3.519 GeV reported by the Selex Collaboration \cite{selex}, but this state needs confirmation \cite{noselexa,noselexb,noselexc}.
\begin{center}
\begin{table}[h]
\caption{Known heavy baryons, with masses in GeV. None of the quantum numbers shown in parentheses have been determined experimentally, but are based on quark model considerations. The states in bold are among those used in the fits of the model to experiment.
\label{charm1}}
\vspace{5mm}
\begin{tabular}{|c|c|c|c|c|c|c|c|c|}
\hline
$\Lambda_c$ & $\Sigma_c$&$\Xi_c$& $\Omega_c$ &$\Lambda_b$&$\Sigma_b$ & $\Xi_b$ &
$\Omega_b$ & $\Xi_{cc}$\\\hline
{\bf 2.285 ($\frac{1}{2}^+$)} & {\bf 2.455 ($\frac{1}{2}^+$)} & {\bf 2.469 ($\frac{1}{2}^+$)} & {\bf 2.698 ($\frac{1}{2}^+$)} & {\bf 5.624 ($\frac{1}{2}^+$)} & {\bf 5.812 ($\frac{1}{2}^+$)} & 5.783 ($\frac{1}{2}^+$) & - & 3.519 ($\frac{1}{2}^+$) \\\hline
{\bf 2.595 ($\frac{1}{2}^-$)}& {\bf 2.518 $(\thlf^+)$} &{\bf 2.577 ($\frac{1}{2}^+$)}& 2.768 ($\frac{3}{2}^+$)& - & {\bf 5.833($\frac{3}{2}^+$)} &- & -& -\\ \hline
2.628 ($\frac{3}{2}^-$)& 2.800 &{\bf 2.647 ($\frac{3}{2}^+$)} &-  &- &- &- &-& -\\ \hline
2.765 & - & 2.789 ($\frac{1}{2}^-$)&-&- &- &- &-& -\\\hline
2.880 ($\frac{5}{2}^\pm$?)& - & 2.817 ($\frac{3}{2}^-$)&-& -& -& -&-& -\\\hline
2.940 & - & 2.980 &-&- &- &- &-& -\\ \hline
-& -& 3.055&-& -&- &- &- & -\\\hline
-& -& 3.080& -&- &- & -&-& -\\\hline
-& -& 3.125& -&- &- &- &-& -\\\hline
\end{tabular}
\end{table}
\end{center}

The baryons containing a single charm quark can be described in terms of SU(3) flavor multiplets, but these represent but a subgroup of the larger SU(4) group that includes all of the baryons containing zero, one, two or three charmed quarks. Furthermore, this multiplet structure is expected to be repeated for every combination of spin and parity, leading to a very rich spectrum of states. One can also construct SU(4) multiplets in which charm is replaced by beauty, as well as place the two sets of SU(4) structures within a larger SU(5) group to account for all the baryons that can be constructed from the five flavors of quark accessible at low to medium energies. It must be understood that the classification of states in SU(4) and SU(5) multiplets serves primarily for enumerating the possible states, as these symmetries are badly broken. Only at the level of the SU(3) ($u,\,\,d,\,\,s)$ and SU(2) $(u,\,\,d)$ subgroups can these symmetries be used in any quantitative way to understand the structure and decays of these states.

\begin{figure}[h]
\includegraphics[width=2.5in]{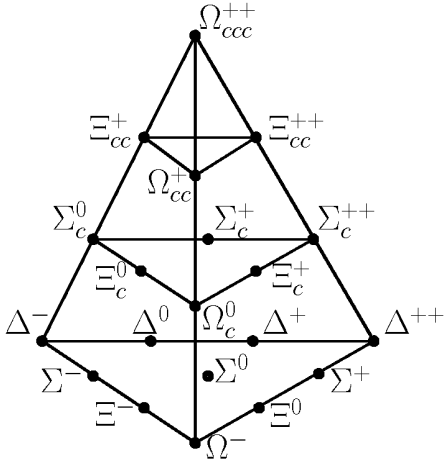}
%\vspace*{-1.0in}
\caption{The symmetric {\bf 20} of SU(4), showing the SU(3) decuplet on the lowest layer.}
\label{fig:multipleta}
\end{figure}

\begin{figure}[h]
\includegraphics[width=2.5in]{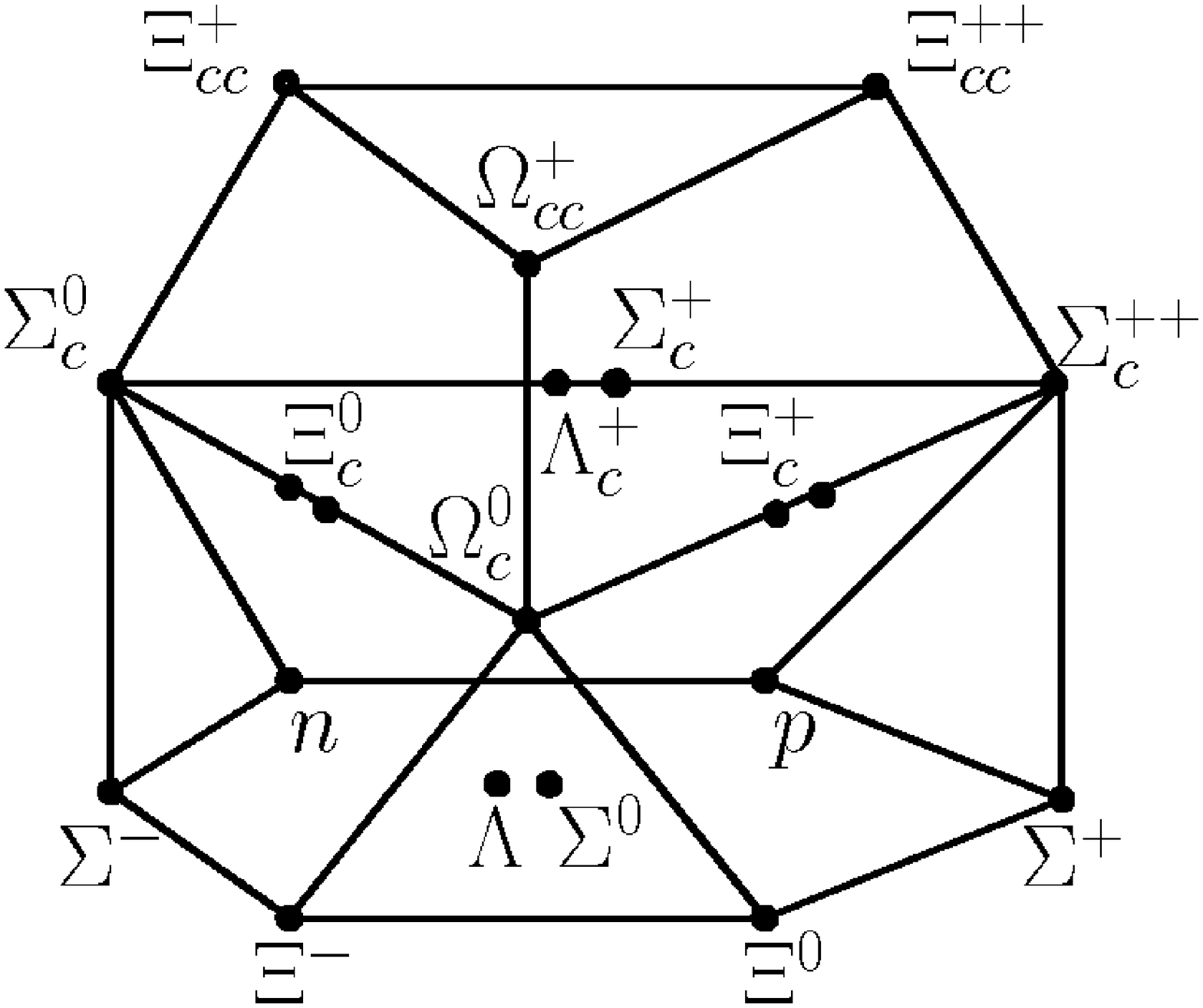}\hskip 0.5in\includegraphics[width=2.0in]{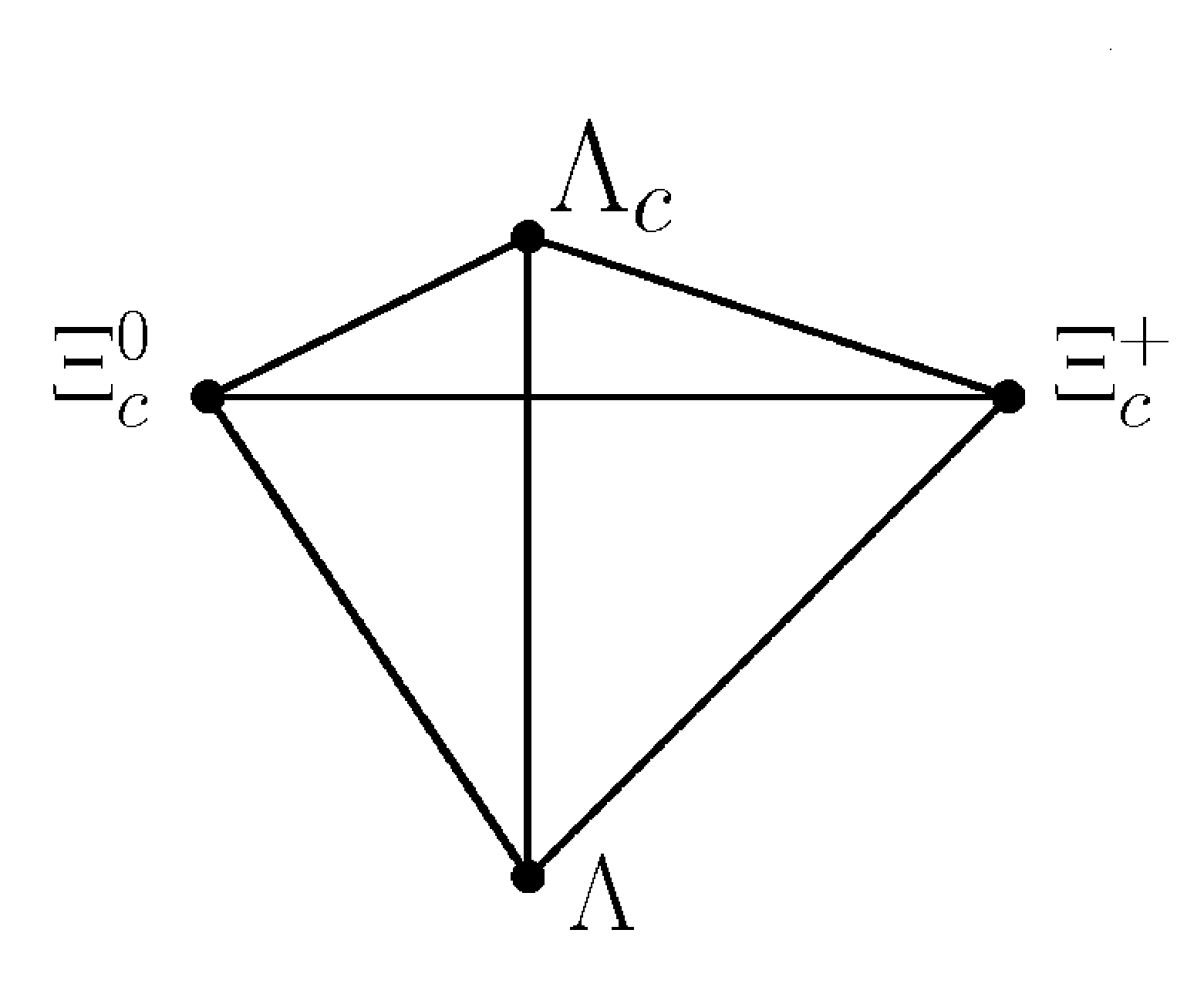}
%\vspace*{-1.3in}
\caption{The mixed-symmetric {\bf 20} (left) and the antisymmetric {\bf 4} (right) of SU(4). The {\bf 20} has the SU(3) octet on the lowest layer, while the {\bf 4} has the SU(3) singlet at the bottom. Note that there two $\Xi_c^+$ and two $\Xi_c^0$ on the middle layer of the {\bf 20}.}
\label{fig:multipletb}
\end{figure}

In flavor SU(3), the baryon multiplets that arise from ${\bf 3\bigotimes 3\bigotimes 3}$ are the well-known decuplet (containing the $\Delta$), two octets (containing the nucleon) and a singlet. The corresponding multiplet structure for SU(4) is ${\bf 4\bigotimes 4\bigotimes 4}= {\bf 20\bigoplus 20\bigoplus 20\bigoplus 4}$. The symmetric {\bf 20} contains the decuplet as a subset, forming the `ground floor' of the weight diagram (shown in Fig. \ref{fig:multipleta}), and all the ground-state baryons in this multiplet have $J^P=\frac{3}{2}^+$. The mixed-symmetric {\bf 20}s (Fig. \ref{fig:multipletb}) contain the octets on the lowest level, and all the ground-state baryons in this multiplet have $J^P=\frac{1}{2}^+$. The ground-floor state of the {\bf 4} (Fig. \ref{fig:multipletb}) is the singlet $\Lambda$ with $J^P=\frac{1}{2}^-$.

%\vspace*{-0.3in}

%\vspace*{-0.3in}

%\vskip 0.3in

Within the flavor SU(3) subgroups, the ground-state heavy baryons containing a single heavy quark belong either to a sextet of flavor symmetric states, or an antitriplet of flavor antisymmetric states, both of which sit on the second layer of the mixed-symmetric {\bf 20} of SU(4) of figure \ref{fig:multipletb}. There is also expected to be a sextet of states with $J^P=\thlf^+$ sitting on the second floor of the symmetric {\bf 20}. The members of the two multiplets of singly-charmed baryons have flavor wave functions
\beqy
\Sigma_c^{++}&=&uuc,\,\,\,\,\Sigma_c^{+}=\frac{1}{\sqrt2}\left(ud+du\right)c,\,\,\,\,\Sigma_c^{0}=ddc\nonumber\\
\Xi_c^{'+}&=&\frac{1}{\sqrt2}\left(us+su\right)c,\,\,\,\,\Xi_c^{'0}=\frac{1}{\sqrt2}\left(ds+sd\right)c,
\nonumber\\
\Omega_c^{0}&=&ssc,
\eeqy
for the sextet and
\beqy
\Lambda_c^{+}=\frac{1}{\sqrt2}\left(ud-du\right)c,\,\,\,\,
\Xi_c^{+}=\frac{1}{\sqrt2}\left(us-su\right)c,\,\,\,\,
\Xi_c^{0}=\frac{1}{\sqrt2}\left(ds-sd\right)c,
\eeqy
for the antitriplet. There is a similar set of flavor wave functions for baryons containing a single $b$ quark. The current flavor multiplet assignments of the lowest-lying charmed baryons is shown in Table \ref{states}.

Isospin symmetry is expected to be very well respected among these states, but SU(3) is more badly broken. It is thus expected that the $\Xi_c$ states observed experimentally will be admixtures of the SU(3) sextet and antitriplet representations. This mixing, induced by many terms in the Hamiltonian, is expected to be small, so that the $\Xi_c$ states at 2.468 and 2.471 GeV, should be predominantly antitriplet with small admixtures of sextet, while the states at 2.576 and 2.578 GeV should be predominantly sextet with small admixtures of antitriplet. This pattern may also occur for the excited states in the spectrum, as well as for any $\Xi_b$ states found. In the case of the latter, the mixing might be expected to be smaller, as HQET arguments suggest that some of the mixing should scale with the inverse of the mass of the heavy quark. However, this mixing cannot be expected to vanish in the heavy quark limit, as SU(3) breaking is independent of this limit.

For the baryons containing two charm or two beauty quarks, the flavor wave functions are 
\beq
\Xi_{cc}^{++}=ccu,\,\,\,\,\Xi_{bb}^{0}=bbu,\,\,\,\,\Omega_{cc}^{+}=ccs,\,\,\,\,\Omega_{bb}^{-}=bbs.
\eeq
When the two heavy quarks are different, there are two ways of constructing their flavor wave functions. One can imagine that the two heavy quarks are members of a (pseudo-)symmetry group, SU(2)$_{bc}$, and that the pair of heavy quarks form either triplet or singlet representations of this group. Two members of this triplet would then be the $\Xi_{cc}$ and $\Xi_{bb}$, with the third member being the state $$\Xi_{bc}^+=\frac{1}{\sqrt2}\left(cb+bc\right)u.$$ The singlet state would then be $$\Xi_{bc}^{'+}=\frac{1}{\sqrt2}\left(cb-bc\right)u.$$ 

Much of the literature on these states treats them essentially in this way. White and Savage \cite{whitesavage} were perhaps the first to argue that the two heavy quarks bind into a color antitriplet source that appears point-like to the remaining light quark. This heavy diquark can then have either spin zero or spin one if the two quarks are different (it can only have spin one if the two quarks have the same flavor). Because the color hyperfine interaction is expected to decrease with increasing heavy-quark mass, the two spin configurations possible in the heavy diquark do not mix at leading order in the heavy quark expansion. 

Alternatively, the flavor wave functions of these two states may be written as $$\Xi_{bc}^{'+}=\frac{1}{\sqrt2}\left(uc+cu\right)b$$ and $$\Xi_{bc}^{+}=\frac{1}{\sqrt2}\left(uc-cu\right)b.$$ The same choices of wave function need to be made when the light quark in the baryon is a strange quark. As with the $\Xi_c$ and $\Xi_b$ states, we expect that there should be mixing between the $\Xi_{bc}$ and $\Xi_{bc}^\prime$ states, whatever representation of the states is chosen, and that this mixing could be large.

\begin{center}
\begin{table}[h]
\caption{Ground-state charmed baryons and their SU(3) assignments. Masses are in GeV. All states have $J^P=\frac{1}{2}^+$.\label{states}}
\vspace{5mm}
\begin{tabular}{|c|c|c|}
\hline
State& Mass& SU(3) assignment\\ \hline
$\Lambda_c^+$&2.285&${\bf \overline{3}}$\\
$\Sigma_c^{++,\,+,\,0}$&2.455&{\bf 6}\\
$\Omega_c^0$&2.698&{\bf 6}\\\hline
$\Xi_c^+$&2.468&${\bf \overline{3}}$\\
$\Xi_c^0$&2.471&${\bf \overline{3}}$\\
$\Xi_c^{'+}$&2.576&{\bf 6}\\
$\Xi_c^{'0}$&2.578&{\bf 6}\\\hline
\end{tabular}
\end{table}
\end{center}
Among the baryons with two or more heavy quarks, the first question to be settled is exactly where do these states lie. The SELEX Collaboration \cite{selex} has published a mass of 3.519 GeV for candidate $\Xi_{cc}$ states, but neither the BaBar \cite{noselexa}, Belle \cite{noselexb} nor Focus \cite{noselexc} Collaborations have found any evidence for these states. Most authors find that the lowest-lying doubly-charmed states are about 100 MeV heavier.

There is a vast literature of theoretical treatments of heavy baryons, including quark models \cite{Copley:1979wj,Stanley:1980fe,Ebert1,Ebert2,Korner,Lyubovitskij:2003pn,Lichtenberg:1996xu,Hwang:1986ee,Lichtenberg:1995kg,faessler,Faessler:2001mr,kiselev02,Rosner:1995yu,Roncaglia:1995az,roncaglia,Albertus:2003sx,Migura:2006ep,Karliner:2007cu,Karliner:2007jp,Karliner:2006ny,Richard:1996za,Richard:1983tc,Martin:1995vk,Martin:1986qt,Hasenfratz:1980ka,SilvestreBrac:1996wp,SilvestreBrac:1996bg,Kalman:1999mk,Chakrabarti:2000rk,CI,Narodetskii:2002ks,narodetskii02,Ebert:2005ip,Ebert:2004sg,Ebert:2002ig,Ebert:1996ec,Albertus:2007xc,Albertus:2006ya,Fleck:1989mb,Cohen:2006jg,martynenko,Gerasyuta:1999pc,Gerasyuta:2007un,Gerasyuta:2008zy,He:2006is,bjorken,Jia:2006gw,Lutz:2003jw,Tong:1999qs,itoh00,Itoh:1992ms,Itoh:1989qi,Itoh:1988xs,Garcilazo:2007eh,Vijande:2006sj,vijande04,gershtein00,Gershtein:1998un,Gershtein:1998sx,Izatt:1981pt,Basdevant:1985ux,Coester:1997ki}, QCD sum rules \cite{Bagan:1992tp,Bagan:1991sc,Wang:2003zp,Wang:2003it,Jin:2001jx,Huang:2000tn,Dai:1995bc,Bagan:1994dy,Bagan:1992za,Liu:2007fg}, treatments in effective field theories \cite{Falk:1996he,Falk:1996qm,Chiladze:1997ev,Mehen:2006vv,Jenkins:1996de,Jenkins:1996rr}, as well as on the lattice \cite{lewis01,mathur02,Mathur:2001id,AliKhan:1999yb,Flynn:2003vz,Na:2007pv,Bowler:1996ws,Alexandrou:1994tc}. We do not attempt to discuss this literature here, but will discuss selected aspects when appropriate later in the manuscript. In the present work, we present the results of a quark-model description of heavy baryons, emphasizing a few aspects that make this work somewhat unique. First, since we explicitly use a quark model with no particular reference to heavy quark symmetries, it is useful to examine how well the model states we obtain for baryons with a single heavy quark reflect the expectations of the heavy quark effective theory. We do this by examining the HQET multiplets expected, and noting which pairs of model states fall into the HQET spin-multiplets. Second, we do not restrict the multiplet structure of the $\Xi_c$ and $\Xi_b$ states, allowing the states from the SU(3) antitriplet and sextet to mix through various terms in the Hamiltonian that we use. We then examine the mixings that result, and compare the masses and wave functions of these states to results we obtain when mixing is not allowed. Third, we examine the double-heavy baryons in the same framework, also exploring the effects of mixing on the spectra of the $\Xi_{bc}$ and $\Omega_{bc}$ states. Fourth, we carry out these analyses not just for the two lowest-lying sets of $J^P$, $\frac{1}{2}^+$ and $\frac{3}{2}^+$, but also for the $\frac{1}{2}^-$, $\frac{3}{2}^-$, $\frac{5}{2}^\pm$ and $\frac{7}{2}^+$ states of the model. Of course, we also attempt to assign model states of particular spin and parity to the experimentally known baryons.

The rest of the manuscript is organized as follows. The model that we use is developed in the next section, while our results for baryons with a single heavy quark are presented in section III. In section IV we present and discuss the results we obtain for baryons containing more than one heavy quark. Our conclusions and outlook are given in section V.

\section{The Model}

Our starting point is a non-relativistic quark model Hamiltonian, similar to that used by Isgur and Karl~\cite{isgurkarla,isgurkarlb,isgurkarlc,isgurkarld,Copley:1979wj}, and applied to a model of the form factors for the semileptonic decays of heavy baryons \cite{Pervin:2005ve,Pervin:2006ie}, and more recently to the hyperons with strangeness -2 and -3 \cite{pr}.

\subsection{Hamiltonian}

The phenomenological Hamiltonian we use takes the form
\beq
\label{hamil}
H=\sum_i K_i +\sum_{i<j}\left( V^{ij}_{\rm conf}+H^{ij}_{\rm hyp}\right) + C_{qqq}.
\eeq
$K_i$ is the kinetic energy of the $i$th quark, with
\beq
K_i= \left( m_i+\frac{p_i^2}{2m_i} \right).
\eeq
The spin independent confining potential consists of linear and Coulomb
components, 
\beq
V^{ij}_{\rm conf}= \sum_{i<j=1}^3\left({br_{ij}\over 2}-
{2\alpha_{\rm Coul}\over3r_{ij}}\right).
\eeq
The spin-dependent part of the potential is written as 
\beq\label{hyp}
H^{ij}_{\rm hyp} =\sum_{i<j=1}^3\left[{2\alpha_{\rm con}\over 3 m_i 
m_j}{8\pi\over 3} \rmb{S}_i\cdot\rmb{S}_j\delta^3(\rmb{r}_{ij})
+ {2\alpha_{\rm ten}\over 3m_i m_j}{1\over {r}^3_{ij}}\left(
{3\rmb{S}_i\cdot\rmb {r}_{ij} \rmb{S}_j\cdot\rmb {r}_{ij}\over {r}^2_{ij}} 
-\rmb{S}_i\cdot\rmb{S}_j\right) \right],
\eeq
which consists of the contact and the tensor terms, with $r_{ij}=\vert\rmb{r}_i-\rmb{r}_j \vert$. The tensor interaction was omitted from the work reported in \cite{Pervin:2005ve,Pervin:2006ie}, but included in the work reported in \cite{pr}. In addition to the interactions described above, we include a simplified spin-orbit potential that takes the form,
\begin{eqnarray}
V_{\rm SO} = \frac{\alpha_{\rm SO}}{\rho^2+\lambda^2} \frac{\rmb{L}\cdot\rmb{S}}{(m_1+m_2+m_3)^2}. 
\end{eqnarray}
In this expression, $L$ is the total orbital angular momentum and $S$ is the total spin of the baryon. We note that this form is not very sensitive to the internal structure of the baryon. It is an {\it ad hoc} form chosen for ease of calculation, and is included in the Hamiltonian to provide an indication of the importance of such a term for the resulting spectrum.

\subsection{Baryon Wave Function}

The total spin of the three spin-$1/2$ quarks can be either $3/2$ or $1/2$.  The
spin wave functions for the maximally stretched state in each case are 
\begin{eqnarray}
\chi_{3/2}^S(+3/2)  &=& |\uparrow\uparrow\uparrow\rangle, \nonumber\\
\chi_{1/2}^\rho(+1/2)  &=& 
\frac{1}{\sqrt{2}}(|\uparrow\downarrow\uparrow\rangle - 
|\downarrow\uparrow\uparrow\rangle), \nonumber\\
\chi_{1/2}^\lambda(+1/2)  &=&- 
\frac{1}{\sqrt{6}}(|\uparrow\downarrow\uparrow
\rangle  + |\downarrow\uparrow\uparrow\rangle -2|
\uparrow\uparrow\downarrow\rangle), \nonumber
\end{eqnarray}
where $S$ labels the state as totally symmetric, while $\lambda/\rho$ denotes
the mixed symmetric states that are symmetric/antisymmetric under the exchange
of quarks $1$ and $2$.

When the masses of the quarks are all different, we choose the Jacobi coordinates to coincide with those for a system in which two of the quarks are identical. This makes it easier to compare results from the two systems. Specifically, we choose, 
\beq
{\bf \rho}= \frac{1}{\sqrt2}({\bf r}_1 - {\bf r}_2),
\eeq
which coincides exactly with the usual definition, and 
\beq
{\bf \lambda} =\sqrt{\frac{2}{3}}\left(\frac{m_1{\bf r}_1 + m_2{\bf r}_2}{m_1+m_2} - 
{\bf r}_3\right).
\eeq
${\bf \rho}$ is the separation of quarks 1 and 2, appropriately normalized, while ${\bf \l}$ is proportional to the separation between the third quark and the center of mass of the 12 pair of quarks.

In our model, a baryon wave function is described in terms of a totally antisymmetric color wave function, multiplying flavor, space and spin wave functions. We use $\phi$ to denote flavor wave functions, $\chi$ for spin, $\psi$ for space, and $\Psi$ for both the spin-space and spin-space-flavor wave functions. The spin-space wave function written for each state is partially determined by its flavor wave function. For flavor wave functions that are (anti)symmetric under exchange of the first two quarks, the spin-space wave function must also be (anti)symmetric under exchange of the first two quarks.  

The spatial wave function for total
${\bf L}={\bf \ell}_\rho+{\bf \ell}_\lambda$ is constructed from a Clebsch-Gordan sum of the wave
functions of the two Jacobi coordinates ${\bf \rho}$ and ${\bf \lambda}$,
and takes the form 
\begin{eqnarray}
\psi_{LMn_{\rho}\ell_{\rho}n_{\lambda}\ell_\lambda}({\bf \rho}, {\bf 
\lambda}) = 
\sum_m\langle LM|\ell_{\rho}m,\ell_\lambda M-m\rangle\psi_{n_\rho \ell_\rho m}
({\bf \rho}) \psi_{n_\lambda \ell_\lambda M-m}({\bf \lambda}).\nonumber
\end{eqnarray}
The spatial and spin wave functions can then be coupled to
give wave functions that are (anti)symmetric in the first two quarks, corresponding to total spin $J$ and parity
$(-1)^{(l_\rho+l_\lambda)}$,
\begin{eqnarray}
\Psi_{JM}&=& \sum_{M_L}\< JM|LM_L, SM-M_L\>\psi_{LM_Ln_{\rho}
\ell_{\rho}n_{\lambda}\ell_\lambda}({\bf \rho}, {\bf 
\lambda})\chi_{S}(M-M_L)\nonumber\\
&\equiv&\left[\psi_{LM_Ln_{\rho}\ell_{\rho}n_{\lambda}\ell_\lambda}({\bf \rho}, {\bf 
\lambda})\chi_{S}(M-M_L)\right]_{J,M}.
\end{eqnarray}
The full wave function for a state $A$ is then built from a linear 
superposition of such components as 
\begin{equation}
\Psi_{A,J^PM}=\phi_A\sum_i \eta_i^A \Psi_{JM}^i.
\end{equation}
In the above, $\phi_A$ is the flavor wave function of the state $A$, and the expansion coefficients $\eta_i^A$ are determined by diagonalizing the Hamiltonian in the
basis of the $\Psi_{JM}$. For this calculation, we limit the expansion in the
last equation to components that satisfy $N\le 2$, where
$N=2(n_\rho+n_\lambda)+\ell_\rho+\ell_\lambda.$  
For states with $J^P=\hlf^+$ in the sextet of SU(3), the spin-space wave functions take the form
\begin{eqnarray}\label{eqmno}
\Psi^{\bf 6}_{\hlf^+M}&=&\left(\left[\vphantom{\sum_i}
\eta_1\psi_{000000}({\bf \rho}, {\bf \lambda})
+\eta_2\psi_{001000}({\bf \rho}, {\bf \lambda})
+\eta_3\psi_{000010}({\bf \rho}, 
{\bf \lambda})\right]\chi_{1/2}^\lambda(M)\right.\nonumber \\
&+&\eta_4\psi_{000101}({\bf \rho}, 
{\bf \lambda})\chi_{1/2}^\rho(M)
+\eta_5\left[\vphantom{\sum_i}\psi_{1M_L0101}({\bf \rho}, {\bf \lambda})
\chi_{1/2}^\rho(M-M_L)\right]_{1/2, M}\nonumber  \\
&+&\left.\eta_6\left[\vphantom{\sum_i}\psi_{2M_L0200}({\bf \rho}, {\bf \lambda})\chi_{3/2}^S(M-M_L)\right]_{1/2, M}
+\eta_7\left[\vphantom{\sum_i}\psi_{2M_L0002}({\bf \rho}, {\bf \lambda})
\chi_{3/2}^S(M-M_L)\right]_{1/2, M}\right),
\end{eqnarray}
while those in the antitriplet are written
\begin{eqnarray}\label{oqmno}
\Psi^{\bf\overline{3}}_{\hlf^+M}&=&\left(\left[\vphantom{\sum_i}
\eta_1\psi_{000000}({\bf \rho}, {\bf \lambda})
+\eta_2\psi_{001000}({\bf \rho}, {\bf \lambda})
+\eta_3\psi_{000010}({\bf \rho}, 
{\bf \lambda})\right]\chi_{1/2}^\rho(M)\right.\nonumber \\
&+&\eta_4\psi_{000101}({\bf \rho}, 
{\bf \lambda})\chi_{1/2}^\lambda(M)
+\eta_5\left[\vphantom{\sum_i}\psi_{1M_L0101}({\bf \rho}, {\bf \lambda})
\chi_{3/2}^S(M-M_L)\right]_{1/2, M}\nonumber  \\
&+&\left.\eta_6\left[\vphantom{\sum_i}\psi_{1M_L0101}({\bf \rho}, {\bf \lambda})\chi_{1/2}^\lambda(M-M_L)\right]_{1/2, M}
+\eta_7\left[\vphantom{\sum_i}\psi_{2M_L0101}({\bf \rho}, {\bf \lambda})
\chi_{3/2}^S(M-M_L)\right]_{1/2, M}\right).
\end{eqnarray}
The wave functions above must be multiplied by the flavor wave function of the state of interest. Note that the spin-space wave functions that multiply the sextet/antitriplet flavor wave functions are valid for any state whose flavor wave function is symmetric/antisymmetric in the first two quarks.  When all three quarks are identical, we use spin-space wave functions that are constructed to be fully symmetric in all three quarks. The wave functions for states of the other spins and parities we consider in this manuscript are shown in Table \ref{wfcomponents}.

In order to examine mixing in the $\Xi_Q$ states in this model, two sets of wave functions are used. In the first set, a spin-space wave function (anti)symmetric in the first two quarks ($us$ or $ds$) is constructed and multiplied by one of the flavor-(anti)symmetric wave functions presented in section I, to create a flavor-space-spin wave function that is symmetric in the first two quarks. In the second set of wave functions that we use, the flavor wave function of a heavy cascade is written as $usQ$, with no (anti)symmetrization in the first two quarks. For states with $J^P=\frac{1}{2}^+$, this flavor wave function multiplies the 14 spin-space wave functions shown in Eqs. (\ref{eqmno}) and (\ref{oqmno}). Mixing between these two sets of states is induced by all of the terms in the Hamiltonian, except for $C_{qqq}$ and the particular form that we use for the spin-orbit interaction. The mixing vanishes in the limit $m_1=m_2$. Some contributions to the mixing are expected to vanish in the limit of an infinitely heavy quark mass, but those arising from the linear and Coulomb parts of the Hamiltonian will not.

For the $\Xi_{bc}^+$, we follow a similar procedure, but examine two different sets of wave functions. In one set, we (anti)symmetrize in the $b$ and $c$ quarks, and examine the mixing between the symmetric and antisymmetric representations. In the second set, we (anti)symmetrize in the $u$ and $c$ quarks, and examine the mixing between these two representations. For the $\Omega_{bc}$, the light quark is replaced with a strange quark, and we follow an analogous procedure.

We construct our wave functions using the harmonic oscillator basis. Each basis wave function takes the well-known form
\beq\label{hoa}
\psi_{nLm} ({\bf r})= \left[\frac{2\,n!}{\left(n + L +\half\right)!}
\right]^{\half} \alpha^{L+\thalf} 
e^{-\frac{\alpha^2r^2}{2}}
L_n^{L+\half}(\alpha^2r^2){\cal Y}_{Lm}({\bf r}),
\eeq
where ${\cal Y}_{Lm}({\bf r})$ is a solid harmonic, and $L_n^{\beta}(x)$ is a generalized Laguerre polynomial. The size parameters $\alpha_\rho$ and $\alpha_\lambda$ appearing in the wave functions are treated as independent variational parameters.

\subsection{Heavy Quark Effective Theory}

One of the many questions of interest is the extent to which the quark model states that we obtain for the baryons containing a single heavy quark respect the dictates of the heavy quark effective theory (HQET). In the HQET description, such a baryon consists of a light component with total spin $j$, coupled to a heavy quark with spin 1/2. The resulting baryon has total angular momentum $J$ that can take the values $J=j\pm 1/2$. The two states with different $J$ are degenerate in the heavy quark limit, with a splitting arising from the chromomagnetic interaction that is suppressed by the mass of the heavy quark. The (almost) degenerate pair of states forms a doublet, and is usually denoted $(J_1, J_2)$, where $J_{1,2}=j\mp 1/2$ \cite{Isgur:1991wq}. In the quark model that we have constructed, the chromomagnetic interaction is suppressed by the mass of the heavy quark, but it is not clear that the states that result have anything to do with the states expected from heavy quark symmetry.

The quark model states we use are constructed in the coupling scheme
\beq
|J^P,L,s_{12}\rangle=\left|\left[\left(\ell_\rho \ell_\lambda\right)_L \left(s_{12} s_3\right)_S\right]_J\right>,
\eeq
where the notation $(ab)_c$ means angular momentum $c$ is formed by
vector addition from angular momenta $a$ and $b$. The parity $P$ is
$(-1)^{\ell_\rho+\ell_\lambda}$, the total spin of the two light
quarks in the baryon is $s_{12}$, and $s_3$ is the spin of the third
quark, taken to be the heavy quark.

The HQET states are assumed to have the coupling scheme
\beq
|J^P,j\rangle=\left|\left\{\left[\left(\ell_\rho \ell_\lambda\right)_L s_{12}\right]_{j} s_3\right\}_J\right>,
\eeq
where $j$ is the total spin of the light component of the baryon, so that $J=j\pm
1/2$. The states of one coupling scheme are linear combinations of the states of the second.
The precise relationship is
\beqy
&&\left|\left\{\left[\left(\ell_\rho \ell_\lambda\right)_L s_{12}\right]_{j} s_3\right\}_J\right>
=(-1)^{1/2+s_{12}+L+J}\sqrt{2j+1}\nonumber\\
&&\times \sum_S\sqrt{2S+1}
\left\{\begin{array}{ccc}1/2&s_{12}&S\\ L& J & j\end{array}\right\}
\left|\left[\left(\ell_\rho \ell_\lambda\right)_L \left(s_{12} s_3\right)_S\right]_J\right>,
\eeqy
where $\left\{\begin{array}{ccc}1/2&s_{12}&S\\ L& J &
j\end{array}\right\}$ is a 6-J symbol. In Appendix \ref{transstates}, we list the heavy baryon multiplets in terms of the quark model quantum numbers that we use.

\subsection{Hamiltonian Parameters and Baryon Spectrum}

In the previous subsections, we introduced the Hamiltonian we use to obtain the baryon spectrum. There are ten free parameters to be determined for
the baryon spectrum: four quark masses ($m_u=m_d$, $m_s$, $m_c$ and $m_b$),
and six parameters of the potential ($\alpha_{\rm con}$, $\alpha_{\rm tens}$, $\alpha_{\rm Coul}$, $b$, $\alpha_{\rm SO}$ and $C_{qqq}$), and these are determined from
a `variational diagonalization' of the Hamiltonian. The variational
parameters are the wave function size parameters $\alpha_\rho$ and
$\alpha_\lambda$ of Eq.~(\ref{hoa}). This variational diagonalization
is accompanied by a fit to the known spectrum, which yields the `best' values for the parameters. Some of the states used in the fit are shown in bold in Table \ref{charm1}. The other states used are the ground-state light hyperons ($\Lambda$, $\Sigma$, $\Xi$ and $\Omega$), their hyperfine partners, where appropriate ($\Sigma^*$, $\Xi^*$), and the nucleon and $\Delta$. These light states serve to provide better constraints on the masses of the light quarks. The results obtained for those states are reported in \cite{pr}.  The
values we obtain for the parameters of the Hamiltonian are shown in
Table~\ref{parameter1}.
\begin{center}
\begin{table}[h]
\caption{Hamiltonian parameters obtained from the fit to a selection of known baryons. \label{parameter1}}
\vspace{5mm}
\begin{tabular}{|c|c|c|c|c|c|c|c|c|c|}
\hline
 $m_\sigma$ & $m_s$  & $m_c$  & $m_b$  & $b$& $\alpha_{\rm Coul}$ 
 &$\alpha_{\rm con}$&$\alpha_{\rm SO}$ &$\alpha_{\rm tens}$ & $C_{qqq}$  \\
 (GeV)&(GeV)&(GeV)&(GeV)&(GeV$^2$)& && (GeV) &&(GeV)\\ \hline
0.2848& 0.5553&  1.8182 & 5.2019 &  0.1540 &$\approx 0.0$&1.0844 &0.9321& -0.2230 & -1.4204\\ \hline
\end{tabular}
\end{table}
\end{center}
These parameters and their implications for baryon spectroscopy have been discussed elsewhere \cite{pr}, but we comment briefly on two aspects. In many of the fits we have obtained, we find that the strength of the Coulomb interaction was consistently small, suggesting that, within this model, that interaction does not play a crucial role. We have also fixed the value of this coupling at 0.1 and 0.2 to investigate its effect on the other parameters and on the spectrum. When this is done, correlations among the parameters mean that they all change but no single parameter changes by more than a few percent. The spectrum also changes, with the masses of states shifting by up to 20 MeV, but with some degradation in the quality of the fit we obtain. Wave function size parameters also change by a few percent. The role of this interaction in the masses of the doubly-heavy baryons will be discussed further, later in the manuscript.

Although the parameter that describes the spin-orbit interaction appears large, the effect of this interaction on the heavy baryons is small. The typical effect on the masses of neglecting this term is a few  ($\lesssim$ 5) MeV. Among the light baryons, the effect is larger, with mass shifts of the order of 20 MeV occuring in some of the negative-parity states.

\section{Baryons with One Heavy Quark}

\subsection{Charmed Baryons with Even Strangeness}

We begin the discussion of our results by examining the predictions of the model for the charmed and beauty baryons with even strangeness. These are the states that clearly belong to the sextet or antitriplet of flavor SU(3). As such, they have been somewhat easier to deal with in models such as these. Our predictions for the spectrum of $\Lambda_c$, $\Sigma_c$ and $\Omega_c$ states are shown in Table \ref{spectrumc1}, while the predictions from a number of other models are shown in Table \ref{baryonspec3}. In Table \ref{spectrumc1}, it can be seen that the known $\Lambda_c$ states are relatively well reproduced by the model, although the ground state is predicted to be somewhat light. The two negative parity excited states are relatively well described. Among the $\Sigma_c$, the model reproduces the two best-known states very well. For the $\Omega_c$, the model prediction is 20 MeV too heavy for the ground state, but this is still within the realm of validity for models like these. The prediction for the excited state is closer to the experimental value. This state was not included in the fit.

In Table \ref{baryonspec3}, it can be seen that, for the most part, the different models are all in agreement with each other. It is not surprsing that the models agree very well for the ground states, but this agreement also extends to the lowest-lying states in spin-parity sectors other than $\hlf^+$. The notable exception to this is seen in the predictions for the negative-parity states in the work of Gerasyuta and co-authors \cite{Gerasyuta:1999pc,Gerasyuta:2007un,Gerasyuta:2008zy}. For these authors, the lowest-lying $\hlf^-$ $\Lambda_c$ is 150 to 200 MeV lighter than in most models (with a mass of 2.4 GeV), and it is the second $\hlf^-$ state that is matched to the experimental state at 2.595 GeV (this is matched with the lowest-lying $\hlf^-$ state in most models). For the higher-lying states in each spin-parity sector, the range in the predictions of the different models is usually larger. One point to note is that, with the exception of the work by Gerasyuta and co-authors, all of the models shown in Table \ref{baryonspec3} indicate that the five lowest-lying negative-parity $\Sigma_c$ all lie within 85 MeV of each other in the model of Garcilazo {\it et al.} \cite{Garcilazo:2007eh}, and within 50 MeV of each other in the other models.

\begin{center}
\begin{table}[h]
\caption{Model predictions for S=0 and S=-2 charmed baryons. All masses are in GeV. The first column identifies the spin and parity of the model state. In this table, only a few of the experimentally known states are assigned to a particular spin and parity, and to a particular model state. Other possible assignments are made in Table \ref{tab:assign}. 
\label{spectrumc1}}
\vspace{5mm}
\begin{tabular}{|l|c|c|c|c|c|c|}
\hline
$J^P$ & \multicolumn{2}{c|}{$\Lambda_c$}& \multicolumn{2}{c|}{$\Sigma_c$}
& \multicolumn{2}{c|}{$\Omega_c$} \\\cline{2-7}
& \hskip 8pt Model \hskip 8pt& Experiment & \hskip 8pt Model\hskip 8pt & Experiment&  \hskip 8pt Model \hskip 8pt &Experiment \\\hline
\multirow{2}{*}{$\frac{1}{2}^+$} & 2.268 &2.285 & 2.455 &2.455 & 2.718 &2.698 \\\cline{2-7}
& 2.791 & - & 2.958 & - & 3.152 & - \\\hline
\multirow{2}{*}{$\frac{3}{2}^+$} & 2.887 & - & 2.519 &2.518 & 2.776& 2.768 \\\cline{2-7}
 & 3.073 & - & 2.995 & - & 3.190 & - \\\hline
\multirow{2}{*}{$\frac{5}{2}^+$} & 2.887 & - & 3.003 & - & 3.196 & - \\\cline{2-7}
 & 3.092 & - & 3.010 & - & 3.203 & - \\\hline
\multirow{2}{*}{$\frac{7}{2}^+$} & 3.128 & - & 3.015 & - & 3.206 & - \\\cline{2-7}
 & - & - & 3.203 & - & 3.327 & - \\\hline
\multirow{2}{*}{$\frac{1}{2}^-$} & 2.625 &2.595 & 2.748& - & 2.977& - \\\cline{2-7}
 & 2.816 &- & 2.768& - & 2.990& - \\\hline
\multirow{2}{*}{$\frac{3}{2}^-$} & 2.636 &2.628 & 2.763 &-  & 2.986& - \\\cline{2-7}
 & 2.830 & - & 2.776 & -  & 2.994& - \\\hline
$\frac{5}{2}^-$ & 2.872 & - & 2.790 & - & 3.014 & -\\\hline
\end{tabular}
\end{table}
\end{center}

\begin{center}
\begin{table}[h]
\caption{Predictions for $\Lambda_c$, $\Sigma_c$ and $\Omega_c$ baryons from a number of quark models
\label{baryonspec3}}

\vspace{5mm}
\begin{tabular}{|c|l|c|l|l|l|l|l|l|}
\hline
Flavor&$J^P$& Expt. &This&\cite {CI}&\cite{Ebert1}&\cite{Migura:2006ep} &\cite{Garcilazo:2007eh}&\cite{Gerasyuta:1999pc,Gerasyuta:2007un,Gerasyuta:2008zy}\\ 
&& Mass&work&&& &&\\ \hline
\multirow{10}{*}{$\Lambda_c$}&\multirow{2}{*}{$\hlf^+$}&2.285&2.268&2.265&2.297&2.272&2.292&2.284\\\cline{3-9}
&&&2.791&2.775&2.772&2.769&2.669&\\\cline{2-9}
&\multirow{2}{*}{$\thlf^+$}&&2.887&2.910&2.874&2.848&2.906&\\ \cline{3-9}
&&&3.073&3.035&3.262&3.100&3.061&\\ \cline{2-9}
&$\fhlf^+$&&2.887&2.910&&&&\\ \cline{2-9}
&\multirow{2}{*}{$\hlf^-$}&2.595&2.625&2.630&2.598&2.594&2.559&2.400\footnotemark[1]\\\cline{3-9}
&&&2.816&2.780&3.017&2.853&2.779&2.635\footnotemark[1]\\\cline{2-9}
&\multirow{2}{*}{$\thlf^-$}&2.628&2.636&2.640&2.628&2.586&2.559&2.625\\\cline{3-9}
&&&2.830&2.840&3.034&2.874&2.779&2.630\\\cline{2-9}
&$\fhlf^-$&&2.872&2.900&&&&2.765\\\hline
\multirow{10}{*}{$\Sigma_c$}&\multirow{2}{*}{$\hlf^+$}&2.455&2.455&2.440&2.439&2.459&2.448&2.458\\\cline{3-9}
&&&2.958&2.890&2.864&2.947&2.793&\\\cline{2-9}
&\multirow{2}{*}{$\thlf^+$}&2.518&2.519&2.495&2.518&2.539&2.505&2.516\\ \cline{3-9}
&&&2.995&2.985&2.912&3.010&2.825&\\ \cline{2-9}
&$\fhlf^+$&&3.003&3.065&&&&\\ \cline{2-9}
&\multirow{2}{*}{$\hlf^-$}&&2.748&2.765&2.795&2.769&2.706&2.700\\\cline{3-9}
&&&2.768&2.770&2.805&2.817&2.791&2.915\\\cline{2-9}
&\multirow{2}{*}{$\thlf^-$}&&2.763&2.770&2.761&2.799&2.706&2.570\\\cline{3-9}
&&&2.776&2.805&2.799&2.815&2.791&2.570\\\cline{2-9}
&$\fhlf^-$&&2.790&2.815&&&&2.740\\\hline
\multirow{10}{*}{$\Omega_c$}&\multirow{2}{*}{$\hlf^+$}&2.698&2.718&&2.698&2.688&2.701&2.806\\\cline{3-9}
&&&3.152&&3.065&3.169&3.044&\\\cline{2-9}
&\multirow{2}{*}{$\thlf^+$}&2.768&2.776&&2.768&2.721&2.759&3.108\\ \cline{3-9}
&&&3.190&&3.119&&3.080&\\ \cline{2-9}
&$\fhlf^+$&&3.196&&&&&\\ \cline{2-9}
&\multirow{2}{*}{$\hlf^-$}&&2.977&&3.020&&2.959&\\\cline{3-9}
&&&2.990&&3.025&&3.029&\\\cline{2-9}
&\multirow{2}{*}{$\thlf^-$}&&2.986&&2.998&&2.959&\\\cline{3-9}
&&&2.994&&3.026&&3.029&\\\cline{2-9}
&$\fhlf^-$&&3.014&&&&&\\\hline
\end{tabular}
\footnotetext[1]{For these authors, the lowest lying $\hlf^-$ state has a mass of 2.4 GeV, and the
experimental state matches their second $\hlf^-$ model state.}
\end{table}
\end{center}

There remain four experimentally known states that have not been assigned in Table \ref{spectrumc1}. These are shown in Table \ref{tab:assign}, along with model states that match the experimental masses relatively closely. The lightest of these is the $\Lambda_c(2765)$. There are two $\Lambda_c$ model states with masses within about 50 MeV of this state, namely the state at 2.791 GeV with $J^P=\frac{1}{2}^+$ (radial excitation), and a state at 2.816 GeV with the same spin but opposite parity. The properties of this experimental state aren't known very well, and it hasn't yet been fully ascertained whether it is a $\Lambda_c$ or a $\Sigma_c$. If it is the latter, then it matches very closely with the model state at 2.768 GeV, with $J^P=\frac{1}{2}^-$. In fact, it matches well with any of the negative parity $\Sigma_c$ states shown in Table \ref{spectrumc1}, as their predicted masses span a narrow band of 45 MeV. 

It was originally suggested that the $\Lambda_c(2880)$ might be a $\frac{1}{2}^-$ state, because of its narrow width \cite{Artuso:2000xy}. A recent analysis of its decays into $\Sigma_c\pi$ by the Belle Collaboration concludes that the angular distribution observed favors $J=\frac{5}{2}$ over $J=\frac{1}{2}$ or $\frac{3}{2}$, but with no determination of the parity \cite{abe}. In our model, states with $J=\frac{5}{2}$ have masses of 2.887 (positive parity) and 2.872 (negative parity) GeV, both excellent matches to this state. Cheng \cite{chenga,cheng} argues that since the ratio $$R=\frac{\Gamma(\Lambda_c(2880)\to\Sigma_c^*\pi)}
{\Gamma(\Lambda_c(2880)\to\Sigma_c\pi)}=22.5\pm 6.2\pm 2.5\%$$ is very different from the value of 1.45 expected from heavy quark symmetry arguments and an assignment of $J^P=\frac{5}{2}^-$, this state must have positive parity. We note here that the predictions of HQET for these decay-rate ratios are always subject to corrections that arise from the $1/m_c$ expansion. Falk and Mehen \cite{falk} have shown that such corrections can lead to large deviations from the expected HQET ratios, in the case of meson decays. We therefore suggest that it might be too early to rule out the possibility of negative parity for this state.

The $\Lambda_c(2940)$ is the heaviest $\Lambda_c$ seen to date. This state is very close to the $D^{*0}p$ threshold, and has been suggested as a candidate molecular bound state \cite{He:2006is}. Our more traditional interpretation offers quark model states with masses of 2.887 GeV, 2.872 GeV and 2.983 GeV as possible matches. If the isospin of this state is in doubt, as has been suggested by Gerasyuta and Matskevich \cite{Gerasyuta:2007un}, the radially excited $\Sigma_c$ with $J^P=\frac{1}{2}^+$ and a mass of 2.958 GeV would provide an excellent match to this state.

The only isovector state in Table \ref{tab:assign} has a mass of 2.800 GeV, which matches well with a number of the negative parity $\Sigma_c$ model states, three of which are shown in the table. It seems unlikely that this state can be assigned to any of the positive parity states, as it is significantly lighter (more than 150 MeV) than any of those model states. We note, however, that models such as this often predict the masses of radial excitations to be too high, especially among baryons composed solely of light quarks. If the isospin of this state is in doubt, then there are a number of $\Lambda_c$ model states of negative parity that are potential matches.

\begin{center}
\begin{table}[h]
\caption{Possible model state and spin-parity assignments for four singly-charmed baryons with even strangeness. The possible assignments are discussed in the text. 
\label{tab:assign}}
\vspace{5mm}
\begin{tabular}{|c|c|c|c|c|}\hline
  &\multicolumn{4}{|c|}{Experimental State} \\\cline{2-5}
\multirow{3}{*}{\begin{tabular}{c}Possible Model States\cr
(flavor, mass (GeV), $J^P$)\end{tabular}}&$\Lambda_c(2765)$ & $\Lambda_c(2880)$ & $\Lambda_c(2940)$&
$\Sigma_c(2800)$\\\cline{2-5}
&$\Lambda_c,\,\,2.791,\,\,\frac{1}{2}^+$ &
$\Lambda_c,\,\,2.887,\,\,\frac{5}{2}^+$ & $\Lambda_c,\,\,2.887,\,\,\frac{3}{2}^+$&
$\Sigma_c,\,\,2.768,\,\,\frac{1}{2}^-$ \\\cline{2-5}
&$\Lambda_c,\,\,2.816,\,\,\frac{1}{2}^-$ &
$\Lambda_c,\,\,2.872,\,\,\frac{5}{2}^-$& $\Lambda_c,\,\,2.872,\,\,\frac{5}{2}^-$& $\Sigma_c,\,\,2.776,\,\,\frac{3}{2}^-$ \\\cline{2-5}
&$\Sigma_c,\,\,2.768,\,\,\frac{1}{2}^-$ & - & $\Lambda_c,\,\,2.983,\,\,\frac{1}{2}^+$ & $\Sigma_c,\,\,2.790,\,\,\frac{5}{2}^-$ \\\hline
\end{tabular}
\end{table}
\end{center}

\subsection{Beauty Baryons with Even Strangeness}

The model predictions for $b$-flavored baryons with S=0 and S=-2 are shown in Table \ref{spectrumb1}, while the results obtained by a number of other authors are shown in Table \ref{baryonspec4}. In the PDG listings, there is only one even-strangeness, $b$-flavored baryon known with any certainty, and that's the $\Lambda_b$. As with the $\Lambda_c$, the model predicts a mass that is too light for this state. There have been recent reports of the masses of the $\Sigma_b$ and $\Sigma_b^*$ \cite{cdf:2007rw}, although these states are not yet in the PDG listings. The present model predicts masses that are larger than the experimental masses for these states, but the splitting between them is well reproduced. At present, there are no experimentally-known $b$ flavored baryons with even strangeness that lack quantum numbers, whether measured or assumed. As with the analagous charmed baryons, the results from different models shown in Table \ref{baryonspec4} agree very well for the lowest lying states in each spin-parity sector. More significant differences among the predictions of the models become apparent for the higher-lying states.

\begin{center}
\begin{table}[h]
\caption{Model predictions for S=0 and S=-2 beauty baryons. All masses are in GeV. The first column identifies the spin and parity of the state. 
\label{spectrumb1}}
\vspace{5mm}
\begin{tabular}{|l|c|c|c|c|c|}
\hline
$J^P$ & \multicolumn{2}{c|}{$\Lambda_b$}& \multicolumn{2}{c|}{$\Sigma_b$}
& \multicolumn{1}{c|}{$\Omega_b$} \\\cline{2-6}
& \hskip 8pt Model \hskip 8pt& Experiment & \hskip 8pt Model\hskip 8pt & Experiment &  \hskip 8pt Model \hskip 8pt \\\hline
\multirow{2}{*}{$\frac{1}{2}^+$} & 5.612 &5.624 & 5.833 &5.812 & 6.081 \\\cline{2-6}
& 6.107 & - & 6.294 & - & 6.472\\\hline
\multirow{2}{*}{$\frac{3}{2}^+$} & 6.181 & - & 5.858 &5.833 & 6.102 \\\cline{2-6}
 & 6.401 & - & 6.308 & - & 6.478 \\\hline
\multirow{2}{*}{$\frac{5}{2}^+$} & 6.183 & -& 6.325 & - & 6.492 \\\cline{2-6}
 & 6.422 & - & 6.328 & - & 6.494 \\\hline
\multirow{2}{*}{$\frac{7}{2}^+$} & 6.433 & - & 6.333 & - & 6.497 \\\cline{2-6}
& - & - & 6.554 & - & 6.667 \\\hline
\multirow{2}{*}{$\frac{1}{2}^-$} & 5.939 & & 6.099& - & 6.301 \\\cline{2-6}
 & 6.180 & - & 6.106& - & 6.312 \\\hline
\multirow{2}{*}{$\frac{3}{2}^-$} & 5.941 & - & 6.101 & - & 6.304 \\\cline{2-6}
 & 6.191 & - & 6.105 & - & 6.311 \\\hline
$\frac{5}{2}^-$ & 6.206 & - & 6.172 & - & 6.311 \\\hline
\end{tabular}
\end{table}
\end{center}

\begin{center}
\begin{table}[h]
\caption{Predictions for $\Lambda_b$, $\Sigma_b$ and $\Omega_b$ baryons from a number of quark models
\label{baryonspec4}}
\vspace{5mm}
\begin{tabular}{|c|l|c|l|l|l|l|}
\hline
Flavor&$J^P$& Expt.&This&\cite {CI}&\cite{Ebert1} &\cite{Garcilazo:2007eh}\\
&&Mass&work&&&\\\hline
\multirow{10}{*}{$\Lambda_b$}&$\hlf^+$&5.624&5.612&5.585&5.622&5.624\\\cline{2-7}
&\multirow{2}{*}{$\thlf^+$}&&6.181&6.145&6.189&6.246\\\cline{3-7}
&&&6.401&&&\\ \cline{2-7}
&$\fhlf^+$&&6.183&6.165&&\\ \cline{2-7}
&\multirow{2}{*}{$\hlf^-$}&&5.939&5.912&5.930&5.890\\\cline{3-7}
&&&6.180&5.780&&5.853\\\cline{2-7}
&\multirow{2}{*}{$\thlf^-$}&&5.941&5.920&5.947&5.890\\\cline{3-7}
&&&6.191&5.840&&5.874\\\cline{2-7}
&$\fhlf^-$&&6.206&6.205&&\\\hline
\multirow{4}{*}{$\Sigma_b$}
&$\hlf^+$&5.812&5.833&5.795&5.805&5.789\\\cline{2-7}
&$\thlf^+$&5.829&5.858&5.805&5.834&5.844\\\cline{2-7}
&$\hlf^-$&&6.099&6.070&6.108&6.039\\\cline{2-7}
&$\thlf^-$&&6.101&6.070&6.076&6.039\\\hline
\multirow{4}{*}{$\Omega_b$}
&$\hlf^+$&&6.081&&6.065&6.037\\\cline{2-7}
&$\thlf^+$&&6.102&&6.088&6.090\\\cline{2-7}
&$\hlf^-$&&6.301&&6.352&6.278\\\cline{2-7}
&$\thlf^-$&&6.304&&6.330&6.278\\\hline
\end{tabular}
\end{table}
\end{center}

\subsection{The $\Xi_c$ Baryons}

From Table \ref{charm1}, there are more $\Xi_c$ states known than any other kind of charmed baryon. This is partly due to the fact that once a resonance signal is extracted from data, the flavor content of these states is easy to identify from the flavor content of the decay products. In the case of the $\Lambda_c$ and $\Sigma_c$, this is usually insufficient to make a definite identification. The flavor content of the final states also helps to identify $\Omega_c$ states.

\begin{center}
\begin{table}[h]
\caption{The $\Xi_c$ spectrum obtained in our model. The first column identifies the state (the spin and parity are the results of the model). The second column shows the masses when no mixing is allowed between the {\bf 6} and ${\bf \overline{3}}$ representations of SU(3), while the third column results when such mixing is allowed. The fourth column shows the experimental masses of the states to which we assign the model states. The fifth column shows the dominant components in the wave function when there is no mixing, while the sixth column shows the same when mixing is allowed. The last column is an estimate of the mixing angle between the two SU(3) representations, as defined in Eq. (\ref{angle}). All masses are in GeV.
\label{spectrumc2}}
\vspace{5mm}
\begin{tabular}{|l|c|c|c|c|c|c|}
\hline
$J^P$& \multicolumn{2}{c|}{Model}& Expt. & \multicolumn{2}{c|}{Dominant Wave Function Components}& $\left|\tan\phi\right|$ \\ \cline{2-3}\cline{5-6}
& Unmixed & Mixed &Mass & Unmixed & Mixed &(Eq. \ref{angle}) \\\hline
\multirow{5}{*}{$\hlf^+$}&2.492&2.466&2.469& $0.9829 |1\rangle_{\bf \overline{3}}+0.1750 |2\rangle_{\bf \overline{3}}$
&$0.9919 |1\rangle_{\bf \overline{3}}+0.0976|3\rangle_{\bf \overline{3}}$&0.0782\\
&&&&
$+0.0399 |3\rangle_{\bf \overline{3}}+0.0411 |4\rangle_{\bf \overline{3}}$&
$+0.0423|2\rangle_{\bf \overline{3}}+0.0438 |1\rangle_{\bf 6}$&\\[+5pt]\cline{2-7}

&2.592&2.594&2.577&$0.9806 |1\rangle_{\bf 6}+0.1709 |3\rangle_{\bf 6}$
&$0.9485 |1\rangle_{\bf 6}-0.2731|2\rangle_{\bf 6}$& 0.0681\\
&&&&
$+0.0877 |2\rangle_{\bf 6}+0.0378 |4\rangle_{\bf 6}$&
$+0.1334|3\rangle_{\bf 6}+0.0603 |4\rangle_{\bf \overline{3}}$&\\\hline

\multirow{5}{*}{$\thlf^+$}&2.650&2.649&2.647&$0.9952 |1\rangle_{\bf 6}+0.0968 |2\rangle_{\bf 6}$
&$0.9915 |1\rangle_{\bf 6}+0.0875|2\rangle_{\bf 6}$&0.0610\\
&&&&&
$+0.0609|1\rangle_{\bf \overline{3}}+0.0744 |3\rangle_{\bf 6}$&\\[+5pt]\cline{2-7}

&2.984&3.012&-& $0.9988|7\rangle_{\bf \overline{3}}-0.0372 |6\rangle_{\bf \overline{3}}$
&$ 0.9783|7\rangle_{\bf \overline{3}}-0.1767|7\rangle_{\bf 6}$&0.1955\\
&&&& $+0.0277|4\rangle_{\bf \overline{3}}$ &
$-0.0574|6\rangle_{\bf \overline{3}}+0.0712 |9\rangle_{\bf 6}$&\\\hline

\multirow{5}{*}{$\fhlf^+$}&2.995&3.004&-& $0.9988 |5\rangle_{\bf \overline{3}}-0.0384 |3\rangle_{\bf \overline{3}}$
&$0.9824 |5\rangle_{\bf \overline{3}}-0.1616|1\rangle_{\bf 6}$&0.1762\\
&&&&$+0.0315|4\rangle_{\bf \overline{3}}$&
$-0.0495|3\rangle_{\bf \overline{3}}+0.0488 |4\rangle_{\bf \overline{3}}$&\\[+5pt]\cline{2-7}

&3.100&3.080&-& $0.9894|5\rangle_{\bf 6}+0.0954 |3\rangle_{\bf 6}$
&$0.9331 |5\rangle_{\bf 6}-0.3179|3\rangle_{\bf \overline{3}}$&0.3500\\
&&&&$+0.0806|4\rangle_{\bf 6}-0.0739|1\rangle_{\bf 6}$&
$-0.0841|5\rangle_{\bf \overline{3}}+0.1200 |3\rangle_{\bf 6}$&\\\hline

\multirow{5}{*}{$\shlf^+$}&3.100&3.094& - &$0.9995 |2\rangle_{\bf 6}+0.0319 |1\rangle_{\bf 6}$
&$0.9683 |1\rangle_{\bf 6}-0.2420|1\rangle_{\bf \overline{3}}$&0.2494\\
& & & & & $+0.0626 |2\rangle_{\bf 6}$& \\[+5pt]\cline{2-7}

&3.216&3.215&-& $|1\rangle_{\bf \overline{3}}$
&$0.9162 |1\rangle_{\bf \overline{3}}-0.3138|1\rangle_{\bf 6}$&0.4374\\
&&&&&
$+0.2492 |2\rangle_{\bf 6}$&\\\hline

\multirow{5}{*}{$\hlf^-$}&2.763&2.773&2.789&$0.9964 |3\rangle_{\bf \overline{3}}-0.0843 |2\rangle_{\bf \overline{3}}$& $0.9849 |3\rangle_{\bf \overline{3}}-0.0961 |2\rangle_{\bf \overline{3}}$&0.1456\\
&&&&&$-0.1117 |1\rangle_{\bf 6}+0.0908 |3\rangle_{\bf 6}$&\\[+5pt]\cline{2-7}

&2.859&2.855&-&$0.9743 |3\rangle_{\bf 6}-0.1534 |1\rangle_{\bf 6}$& $0.9386 |3\rangle_{\bf 6}-0.1871 |2\rangle_{\bf 6}$&0.2846\\
&&&&$-0.1649 |2\rangle_{\bf 6}$&$-0.2384 |2\rangle_{\bf \overline{3}}-0.1217 |3\rangle_{\bf \overline{3}}$&\\\hline

\multirow{5}{*}{$\thlf^-$}&2.784&2.783&2.817&$0.9953 |3\rangle_{\bf \overline{3}}-0.0973 |2\rangle_{\bf \overline{3}}$& $0.9848 |3\rangle_{\bf \overline{3}}-0.0966 |2\rangle_{\bf \overline{3}}$&0.1462\\
&&&&&$-0.1125 |1\rangle_{\bf 6}+0.0909 |3\rangle_{\bf 6}$&\\[+5pt]\cline{2-7}

&2.871&2.866&-& $0.9828 |3\rangle_{\bf 6}-0.1576 |1\rangle_{\bf 6}$& $0.9510 |3\rangle_{\bf 6}-0.1105 |2\rangle_{\bf 6}$&0.2822\\
&&&&$+0.0959 |2\rangle_{\bf 6}$&$-0.2406 |2\rangle_{\bf \overline{3}}-0.1228 |3\rangle_{\bf \overline{3}}$&\\\hline

\multirow{3}{*}{$\fhlf^-$}&2.905&2.895& - &$|1\rangle_{\bf 6}$& $0.9763 |1\rangle_{\bf 6}-0.2165 |1\rangle_{\bf \overline{3}}$&0.2218\\[+5pt]\cline{2-7}

&2.984&2.989& - &$ |1\rangle_{\bf \overline{3}}$& $ 0.9763|1\rangle_{\bf \overline{3}}+0.2165|1\rangle_{\bf 6}$&0.2218\\[+5pt]\hline

\end{tabular}
\end{table}
\end{center}

Our model results for the $\Xi_c$ states are shown in Table~\ref{spectrumc2}, and those of a number of other authors are shown in Table \ref{baryonspec5}. As has been noted for other flavor sectors, the different models agree quite well in their predictions for the lowest-lying states in each spin-parity sector, but predictions for the higher-lying states show a bit more disagreement. Nevertheless, for the states shown, all model predictions are within 100 MeV of each other. 

The first column in Table \ref{spectrumc2} shows the angular momentum and parity of the quark model states. Column two shows the model masses obtained when the states are treated as purely sextet or antitriplet flavor states, while column three shows the masses that result when the two flavor multiplets are allowed to mix. The fourth column shows the experimental mass of the state to which we assign the quark model state.  Column five shows the dominant components of the wave function in the unmixed case, while column six shows the largest contributions to the mixed wave functions. Column seven shows the tangent of the mixing angle as defined in Eq. (\ref{angle}). 

\begin{center}
\begin{table}[h]
\caption{Predictions for $\Xi_c$ baryons from a number of quark models
\label{baryonspec5}}
\vspace{5mm}
\begin{tabular}{|l|c|l|l|l|l|}
\hline
$J^P$& Expt. &This&\cite{Ebert1} &\cite {Migura:2006ep}&\cite{Garcilazo:2007eh}\\
& mass&work& &&\\ \hline

\multirow{2}{*}{$\hlf^+$}&2.469&2.466&2.481&2.469&2.496\\\cline{2-6}
&2.577&2.594&2.578&2.595&2.574\\\hline
\multirow{2}{*}{$\thlf^+$}&2.647&2.649&2.654&2.651&2.633\\ \cline{2-6}
&&3.012&3.030&&2.951\\ \hline
$\fhlf^+$&&3.004&&&\\ \hline
\multirow{2}{*}{$\hlf^-$}&2.789&2.773&2.801&2.769&2.749\\\cline{2-6}
&&2.855&2.928&&2.829\\\hline
\multirow{2}{*}{$\thlf^-$}&2.817&2.783&2.820&2.771&2.749\\\cline{2-6}
&&2.866&2.900&&2.829\\\hline
$\fhlf^-$&&2.989&&&\\\hline
\end{tabular}
\end{table}
\end{center}

As can be seen from the table, the model is reasonable successful in describing the  $\Xi_c$ states with assigned quantum numbers. For those states whose quantum numbers are not yet assigned or measured, Table \ref{tab:xi_assign} shows the model states that match most closely in mass to the experimental states. As there is little experimental information to constrain the assignment of possible model states to a particular experimental state, we impose the condition that model states should not be highly excited ones. For instance, the fifth and sixth $\frac{1}{2}^-$ and $\frac{3}{2}^-$ states have masses that are close to the $\Xi_c(2980)$, but we do not consider them at this point. We will discuss these assignments in more detail below.

It appears odd that, when mixing is included in the model, the mass of lowest state in some $J^P$ sectors, such as the $\hlf^-$ and $\fhlf^+$ sectors, increases. This is contrary to what is expected: mixing lowers the mass of the lowest-lying state in any sector. In our model, this arises because we diagonalize the Hamiltonian anew when we consider mixing. Because the wave functions are determined in a variational way, the wave function size parameter sets obtained for the two unmixed flavor sectors for a particular $J^P$ are different from each other, and are also different from those obtained when mixing is included. It is this change in the size parameters, $\alpha_\rho$ and $\alpha_\lambda$, that is responsible for the increases in mass when mixing is included.

One key result of the model in this sector is the mixing between the flavor sextet and antitriplet states. From previous studies, this mixing, usually characterized in terms of a mixing angle, is small. However, to the best of our knowledge, it has only been explored for the ground state heavy cascades. In the present model, mixing arises from most of the terms in the Hamiltonian. For three quarks of different masses, we note that
\beq
{\bf r}_{13}=\sqrt{2}\left(\frac{m_2}{m_1+m_2}{\bf \rho}+\frac{\sqrt{3}}{2}{\bf \lambda}\right),\,\,\,\,
{\bf r}_{23}=\sqrt{2}\left(-\frac{m_1}{m_1+m_2}{\bf \rho}+\frac{\sqrt{3}}{2}{\bf \lambda}\right).
\eeq
The mass dependence in $r_{23}$ and $r_{13}$ provides one contribution to mixing between sextet and antitriplet wave function components that arises from the linear and Coulomb terms in the Hamiltonian, even in the absence of spin-dependent forces.

Recall that our wave functions are defined in terms of a number of components, such as
\beq
\left|\hlf^+\right\rangle_f=\sum_{i=1}^7\eta_i^f \left|i,\hlf^+\right\rangle_f,
\eeq
for example, where $f$ denotes the flavor multiplet, the components $\left|i,\hlf^+\right\rangle_f$ are given in Appendix \ref{components}, and the $\eta_i^f$ are the expansion coefficients that result from the variational diagonalization of the Hamiltonian. With such wave functions, a `mixing matrix' can be defined, but there is no simple way to define a `mixing angle' between the sextet and antitriplet components of a state, particularly as the mixing interactions are not treated perturbatively. For this calculation, we choose an operational definition, or prescription, as follows. Each state is written in terms of sextet and antitriplet components. For a state that is predominantly antitriplet, we define a `mixing angle' $\phi$ as
\beq\label{angle}
\tan^2{\phi}\equiv\frac{\sum_{i=1}^{N_6} {\eta_i^6}^2}{\sum_{i=1}^{N_{\overline{3}}} 
{\eta_i^{\overline{3}}}^2},
\eeq
where $\eta_i^f$ are the expansion coefficients corresponding to flavor $f$ in the wave function, and $N_f$ is the number of components of flavor $f$ in the wave function. For most cases, $N_6=N_{\overline{3}}$. This definition of the angle provides information on the relative sizes of the sextet and antitriplet components of the wave function. This angle is as defined in Eq. (\ref{angle}) for antitriplet states, and the reciprocal of this definition for sextet states. For all of the states shown in Table \ref{spectrumc2}, the wave function is dominated by a single component, even when mixing is included. The mixing angles are small for most of the states, but for some states they become quite large, with $|\tan{\phi}|$ approaching 0.5. For the two lowest lying states, the value of $\tan{\phi}$ translates into a mixing angle of less than 4.5$^\circ$, consistent with the results of Franklin \cite{franklinprime}.

In Table \ref{charm1}, there are a number of experimentally observed $\Xi_c$ states that have no spin-parity assignments, and these states have been omitted from Table \ref{spectrumc2}. In Table \ref{tab:xi_assign} we present these states along with a number of model states that have masses that are close to the experimental ones. In this table, we treat the states at 3.055 GeV and 3.080 GeV together as they are close in mass, and therefore have a number of candidate model states in common. It is clear that each of the new $\Xi_c$ baryons can be identified with any of a number of quark model states. Analysis of the decays of these states, both experimentally and within the context of a model such as this, is necessary for identifying which quark model state best matches which experimental state.

\begin{center}
\begin{table}[h]
\caption{Possible model state and spin-parity assignments for four singly-charmed baryons with odd strangeness. The possible assignments are discussed in the text. 
\label{tab:xi_assign}}
\vspace{5mm}
\begin{tabular}{|c|c|c|c|}\hline
  &\multicolumn{3}{|c|}{Experimental State} \\\cline{2-4}
\multirow{7}{*}{\begin{tabular}{c}Possible Model States\cr
(flavor, mass (GeV), $J^P$)\end{tabular}}&$\Xi_c(2980)$ & $\Xi_c(3055)$, $\Xi_c(3080)$&
$\Xi_c(3125)$\\\cline{2-4}
&$2.924,\,\,\frac{1}{2}^+$ &$3.012,\,\,\frac{3}{2}^+$ & $3.136,\,\,\frac{1}{2}^+$ \\\cline{2-4}
&$3.012,\,\,\frac{3}{2}^+$ &$3.075,\,\,\frac{3}{2}^+$& $3.094,\,\,\frac{7}{2}^+$ \\\cline{2-4}
&$3.004,\,\,\frac{5}{2}^+$ & $3.080,\,\,\frac{3}{2}^+$& - \\\cline{2-4}

&$2.989,\,\,\frac{5}{2}^-$ & $3.092,\,\,\frac{3}{2}^+$& -\\\cline{2-4}
& - &$3.004,\,\,\frac{5}{2}^+$&  - \\\cline{2-4}
&- & $3.080,\,\,\frac{5}{2}^+$& - \\\cline{2-4}
& -& $3.092,\,\,\frac{5}{2}^+$& - \\\cline{2-4}
& -& $3.094,\,\,\frac{7}{2}^+$& - \\\hline
\end{tabular}
\end{table}
\end{center}

\subsection{The $\Xi_b$ Baryons}

Our results for the $\Xi_b$ states are shown in Table~\ref{spectrumb2}, and those of a few other authors are shown in Table \ref{baryonspec6}. In Table \ref{spectrumb2}, the columns are as in Table \ref{spectrumc2}. While this manuscript was being prepared, the CDF \cite{cdf:2007un} and D0 \cite{d0:2007ub} Collaborations reported results on the mass of the first $\Xi_b$ state observed, noting that it is the first observed baryon formed of quarks from all three families. We have included that result in the table below, noting that this state was not included in our fits to the baryon spectrum. Our result for this state is somewhat higher than the experimental results, but in quite good agreement with them.

The values of $|\tan{\phi}|$ shown in Table \ref{spectrumb2} are all smaller than the corresponding values shown in Table \ref{spectrumc2}. We have also calculated a spectrum of $\Xi_b$ states in which the mass of the $b$ quark was 45.5 GeV. The mixing angles obtained in that spectrum are not very different from those shown in Table \ref{spectrumb2}, and some of them are slightly larger, indicating that some of the mixing is not getting smaller as the quark mass gets larger.

\begin{center}
\begin{table}[h]
\caption{The $\Xi_b$ spectrum obtained in our quark model. The key to the columns is as in Table \ref{spectrumc2}. All masses are in GeV.
\label{spectrumb2}}
\vspace{5mm}
\begin{tabular}{|l|c|c|c|c|c|c|}
\hline
$J^P$& \multicolumn{2}{c|}{Model}& Expt. &\multicolumn{2}{c|}{Dominant Wave Function Components}& $\left|\tan\phi\right|$ \\ \cline{2-3}\cline{5-6}
& Unmixed & Mixed & Mass& Unmixed & Mixed & (Eq. \ref{angle})\\\hline
\multirow{5}{*}{$\hlf^+$} &5.844&5.806&5.774 (D0)& $0.9700 |1\rangle_{\bf \overline{3}}+0.2012 |2\rangle_{\bf \overline{3}}$
&$0.9913 |1\rangle_{\bf \overline{3}}+0.1212|3\rangle_{\bf \overline{3}}$&0.0350\\
& &&5.795 (CDF) &
$+0.1356 |3\rangle_{\bf \overline{3}}+0.0135 |4\rangle_{\bf \overline{3}}$&
$+0.0358|2\rangle_{\bf \overline{3}}+0.0330 |4\rangle_{\bf 6}$&\\[+5pt]\cline{2-7}

& 5.958&5.970&- &$0.9978 |1\rangle_{\bf 6}+0.0523 |3\rangle_{\bf 6}$
&$0.9452 |1\rangle_{\bf 6}-0.3023|2\rangle_{\bf 6}$&0.0548\\
&&&&
$+0.0359 |2\rangle_{\bf 6}+0.0199 |4\rangle_{\bf 6}$&
$+0.1083|3\rangle_{\bf 6}+0.0528 |4\rangle_{\bf \overline{3}}$&\\\hline

\multirow{5}{*}{$\thlf^+$}& 5.982&5.980& - &$0.9997 |1\rangle_{\bf 6}+0.0183 |2\rangle_{\bf 6}$
&$0.9947 |1\rangle_{\bf 6}+0.0731|2\rangle_{\bf 6}$&0.0598\\
&&&&&$+0.0597|1\rangle_{\bf \overline{3}}+0.0395 |3\rangle_{\bf 6}$&\\[+5pt]\cline{2-7}

& 6.294&6.311&- &$0.9996 |7\rangle_{\bf \overline{3}}+0.0243 |4\rangle_{\bf \overline{3}}$
&$0.9887 |7\rangle_{\bf \overline{3}}-0.1411|7\rangle_{\bf 6}$&0.1452\\
&&&&$-0.0155|6\rangle_{\bf \overline{3}}$&
$+0.0371|4\rangle_{\bf \overline{3}}+0.0264 |9\rangle_{\bf 6}$&\\\hline

\multirow{5}{*}{$\fhlf^+$}& 6.333&6.300&- &$0.9994 |5\rangle_{\bf \overline{3}}+0.0307 |4\rangle_{\bf \overline{3}}$
&$0.9907 |5\rangle_{\bf \overline{3}}-0.1277|1\rangle_{\bf 6}$&0.1312\\
&&&&$-0.0150|3\rangle_{\bf \overline{3}}$&
$+0.0336|4\rangle_{\bf \overline{3}}+0.0243 |5\rangle_{\bf 6}$&\\[+5pt]\cline{2-7}

& 6.402&6.393&- &$0.9953 |5\rangle_{\bf 6}+0.0769 |4\rangle_{\bf 6}$
&$0.9617 |5\rangle_{\bf 6}-0.2518|3\rangle_{\bf \overline{3}}$&0.2635\\
&&&&$+0.0514|3\rangle_{\bf 6}-0.0304 |1\rangle_{\bf 6}$&
$+0.686|4\rangle_{\bf 6}+0.0704 |3\rangle_{\bf 6}$&\\\hline

\multirow{5}{*}{$\shlf^+$}& 6.405&6.395&- &$0.9996 |2\rangle_{\bf 6}+0.0301 |1\rangle_{\bf 6}$
&$0.9718 |2\rangle_{\bf 6}-0.2302|1\rangle_{\bf \overline{3}}$&0.2365\\
&&&&&
$+0.0514|1\rangle_{\bf 6}$&\\[+5pt]\cline{2-7}

& 6.524&6.517&- &$|1\rangle_{\bf \overline{3}}$
&$0.9558 |1\rangle_{\bf \overline{3}}+0.2357|2\rangle_{\bf 6}$&0.3078\\
&&&&&
$-0.1760|2\rangle_{\bf 6}$&\\\hline

\multirow{5}{*}{$\hlf^-$}& 6.108&6.090&- &$0.9996 |3\rangle_{\bf \overline{3}}-0.0292 |2\rangle_{\bf \overline{3}}$& $0.9953 |3\rangle_{\bf \overline{3}}-0.0866 |1\rangle_{\bf 6}$&0.0917\\
&&&&&$+0.0288 |3\rangle_{\bf 6}-0.0313 |2\rangle_{\bf \overline{3}}$&\\[+5pt]\cline{2-7}

& 6.192&6.188&- &$0.9903 |3\rangle_{\bf 6}-0.1305 |2\rangle_{\bf 6}$& $0.9706 |3\rangle_{\bf 6}-0.1537 |2\rangle_{\bf 6}$&0.1837\\
&&&&$-0.0479 |1\rangle_{\bf 6}$&$-0.1728 |2\rangle_{\bf \overline{3}}-0.0414 |1\rangle_{\bf 6}$&\\\hline

\multirow{5}{*}{$\thlf^-$}& 6.110&6.093&- &$0.9996 |3\rangle_{\bf \overline{3}}-0.0292 |2\rangle_{\bf \overline{3}}$& $0.9953 |3\rangle_{\bf \overline{3}}-0.0867 |1\rangle_{\bf 6}$&0.0918\\
&&&&&$-0.0313 |2\rangle_{\bf \overline{3}}+0.0288 |3\rangle_{\bf 6}$&\\[+5pt]\cline{2-7}

& 6.194&6.190&- &$0.9968 |3\rangle_{\bf 6}+0.0639 |2\rangle_{\bf 6}$& $0.9806 |3\rangle_{\bf 6}-0.1726 |2\rangle_{\bf \overline{3}}$&0.1801\\
&&&&$-0.0484 |1\rangle_{\bf 6}$&$+0.0728 |2\rangle_{\bf 6}-0.0421 |1\rangle_{\bf 6}$&\\\hline

\multirow{3}{*}{$\fhlf^-$}& 6.204&6.201&- &$|1\rangle_{\bf 6}$& $0.9864 |1\rangle_{\bf 6}-0.1642 |1\rangle_{\bf \overline{3}}$&0.1665\\[+5pt]\cline{2-7}

& 6.312&6.313&- &$|1\rangle_{\bf \overline{3}}$& $0.9864 |1\rangle_{\bf 6}+0.1642 |1\rangle_{\bf \overline{3}}$&0.1665\\[+5pt]\hline
\end{tabular}
\end{table}
\end{center}

\begin{center}
\begin{table}[h]
\caption{Predictions for $\Xi_c$ baryons from a number of quark models.
\label{baryonspec6}}
\vspace{5mm}
\begin{tabular}{|l|c|l|l|l|}
\hline
$J^P$&Expt. &This&\cite{Ebert1} &\cite{Garcilazo:2007eh}\\
&mass&work&&\\\hline
$\hlf^+$&5.780&5.806&5.812&5.825\\\hline
$\thlf^+$&&5.980&5.963&5.967\\\hline
$\hlf^-$&&6.090&6.119&6.076\\\hline
$\thlf^-$&&6.093&6.130&6.076\\\hline
\end{tabular}
\end{table}
\end{center}

\subsection{HQET and Spin Multiplets}

The heavy quark effective theory predicts that baryon states containing a single heavy quark should fall into almost degenerate multiplets. If the light component of the baryon has total angular momentum $j$, inclusion of the spin of the heavy quark means that two states are possible, with total angular momentum $J=j\pm 1/2$ (usually denoted $((j-1/2)^P,(j+1/2)^P)$). These two states have the same parity as the light component and, because the part of the hyperfine interaction that involves the heavy quark is suppressed by the mass of the heavy quark, these two states should be degenerate in mass in the limit when the heavy quark in the baryon becomes infinitely massive. The exception occurs when the light component of the baryon has total angular momentum zero, in which case the spin of the baryon can only be 1/2.

Among the sextet baryons, the expansion up to the $N=2$ harmonic oscillator band provides wave functions and masses for seven states with $J^P=\frac{1}{2}^+$, nine states with $J^P=\frac{3}{2}^+$, five states with $J^P=\frac{5}{2}^+$, and two states with $J^P=\frac{7}{2}^+$. Among the baryons with negative parity, the expansion gives three states with $J^P=\frac{1}{2}^-$, three with $J^P=\frac{3}{2}^-$, and a single state with $J^P=\frac{5}{2}^-$. Among the antitriplet baryons, the counting of the negative parity states remains the same. For the positive parity states, there are seven states with $J^P=\frac{1}{2}^+$, seven with $J^P=\frac{3}{2}^+$, five with $J^P=\frac{5}{2}^+$, and a single state with $J^P=\frac{7}{2}^+$.

In order to place these states into HQET multiplets, we must assume that the counting described above is `complete', meaning that, for instance, the states in the $N=2$ band  with $J^P=\frac{7}{2}^+$ can be HQET partners only with states also in the $N=2$ band. This makes some sense intuitively, as states from higher bands (in this case, it would have to be the $N=4$ band) should be considerably heavier. If we consider an expansion in the spin-space wave function that goes beyond the $N=2$ band, then the statement would refer to states whose wave function components lie predominantly in the $N=2$ (or lower) band. 

For the sextet states, modulo the argument about harmonic oscillator bands, there is only one possible way to account for the 23 states with positive parity and the seven states with negative parity. For positive parity, there must be two $(\frac{5}{2}^+,\frac{7}{2}^+)$ doublets, three $(\frac{3}{2}^+,\frac{5}{2}^+)$ doublets, six $(\frac{1}{2}^+,\frac{3}{2}^+)$, and a lone $(\frac{1}{2}^+)$ singlet. Among the negative parity states, there must be one $(\frac{3}{2}^-,\frac{5}{2}^-)$ doublet, two $(\frac{1}{2}^-,\frac{3}{2}^-)$ doublets, and one $(\frac{1}{2}^-)$ singlet. For the states that fall into the antitriplet, the multiplets with negative parity are the same as for the sextet states. Among the states with positive parity, there must be a single $(\frac{5}{2}^+,\frac{7}{2}^+)$ doublet, four $(\frac{3}{2}^+,\frac{5}{2}^+)$ doublets, three $(\frac{1}{2}^+,\frac{3}{2}^+)$ doublets, and four $(\frac{1}{2}^+)$ singlets.

There are many almost-degenerate pairs of states in our model spectrum that might appear to constitute the multiplets expected from HQET. Proper identification of the spin doublets requires examination of the structure of the wave functions of the states. All of the wave functions are rewritten in terms of the HQET states shown in Appendix \ref{transstates}, and a state is identified with one of the HQET states if the expansion coefficient corresponding to that HQET state is greater than 0.91 (corresponding to a `mixing angle' of 25$^\circ$). A doublet is identified if both members meet this criterion.  Among the antitriplet states, the spin multiplets that we have identified in this way are shown in Table \ref{antitriplet}, while the corresponding states for the sextets are shown in Table \ref{sextet}. 

In discussing these multiplets, we need to discuss two different kinds of mixing among the states. Consider the states with $J^P=\thlf^+$, for instance. These states can belong either to $(\hlf,\thlf)$ doublets or $(\thlf, \fhlf)$ doublets. If the spin $\thlf$ states in the $(\thlf, \fhlf)$ multiplets mix with each other, we refer to this as `intra-multiplet mixing', but if they mix with the spin $\thlf$ states of the $(\hlf,\thlf)$ multiplets, we define this as `cross-multiplet mixing'. We have also generated a spectrum of states assuming that the mass of the heavy quark is 45.5 GeV, as this will provide us with some insight on how well the quark model approaches the expectations of HQET.

Among the flavor antitriplet states, three of the positive parity spin-singlets are among the easiest HQET states to identify. The other spin singlet, the second of Eq. (\ref{spinsinglet}), has a slightly more complicated structure, but is nevertheless identifiable. It is clear that we have not grouped all of the antitriplet states in spin multiplets. For the states not shown, in many cases one member of the spin multiplet could be identified, but the second member could not. In other cases, neither member could be clearly identified. For instance, in each flavor sector, there should be three $(\hlf^+,\thlf^+)$ multiplets, but the spin-$\hlf$ states all show strong intra-multiplet mixing except for one multiplet in the case of the $\Lambda_b$. The spin-$\frac{1}{2}$ states in these multiplets show very little cross-multiplet mixing, but there is significant cross-multiplet mixing in some of the spin-$\thlf$ states. In addition, the cross-multiplet mixing that exists decreases when the mass of the heavy quark is increased, and this is seen in the fact that we are able to identify one more $(\thlf^+,\fhlf^+)$ multiplet in the case of the $\Lambda_b$ baryons. Nevertheless, the strong intra-multiplet mixing in the positive parity sector persists even with a very large mass for the heavy quark. This means that these states in the $c$ or $b$ sector, if found, may exhibit behavior that departs from the predictions of HQET. There is also strong cross-multiplet mixing in the negative parity states of the SU(3) antitriplets. Among the heavy cascades, the additional mixing between flavor antitriplet and flavor sextet states means that fewer states can be identified as HQET states.

For the sextet states, some of the trends are the same. Among the positive parity states, there are multiplets that cannot be identified primarily because of intra-multiplet mixing, but there are also a few where strong cross-multiplet mixing manifests itself, and this persists even when the mass of the heavy quark is made very large. Indications are that these states are tending to the HQET states quite slowly as the mass of the heavy quark is increased. It must be noted that these states are highly excited states, and it is probable that our truncated expansion may be running into the boundaries of the reliability phase space.

\begin{center}
\begin{table}[h]
\caption{Model predictions for antitriplet HQET spin multiplets.
\label{antitriplet}}
\vspace{5mm}
\begin{tabular}{|c|c|c|c|c|}
\hline
$(j-1/2^P,j+1/2^P)$ & $\Lambda_c$& $\Lambda_b$ & $\Xi_c$ & $\Xi_b$\\\hline
\multirow{4}{*}{$\left(\frac{1}{2}^+\right)$} & 2.268 &5.612 & 2.466 & 5.806  \\\cline{2-5}
 & 2.791 &6.107 & 2.924 &6.230  \\\cline{2-5}
 & 2.983 &6.338 & 3.183 &6.547  \\\cline{2-5}
 & 3.154 &6.499 & - &6.719  \\\hline
$\left(\frac{1}{2}^+,\frac{3}{2}^+\right)$ & - & (6.423, 6.401) & - & - \\\hline
\multirow{4}{*}{$\left(\frac{3}{2}^+,\frac{5}{2}^+\right)$} & (2.887, 2.887) &(6.181, 6.183) & (3.012, 3.004) & (6.311, 6.300) \\\cline{2-5}
 & (3.120, 3.125) &(6.431, 6.434) & - & (6.528, 6.529) \\\cline{2-5}
 & (3.194, 3.194) &(6.449, 6.450) & - & - \\\cline{2-5}
& - &(6.549, 6.549) & - & - \\\hline
$\left(\frac{5}{2}^+,\frac{7}{2}^+\right)$ & (3.092, 3.128) & (6.422, 6.433) & - & - \\\hline
$\left(\frac{1}{2}^-\right)$ & - & 6.180 & - & - \\\hline
\multirow{2}{*}{$\left(\frac{1}{2}^-,\frac{3}{2}^-\right)$} & (2.625, 2.636) & (5.939, 5.941) & (2.773, 2.783) & (6.090, 6.093)\\\cline{2-5}
& - & (6.206, 6.211) & - & - \\\hline
$\left(\frac{3}{2}^-,\frac{5}{2}^-\right)$ & (2.830, 2.872) & (6.191, 6.206) & - & - \\\hline
\end{tabular}
\end{table}
\end{center}

\begin{center}
\begin{table}[h]
\caption{Model predictions for sextet HQET spin multiplets.
\label{sextet}}
\vspace{5mm}
\begin{tabular}{|c|c|c|c|c|c|c|}
\hline
$(j-1/2^P,j+1/2^P)$ & $\Sigma_c$& $\Sigma_b$ & $\Xi_c$ & $\Xi_b$& $\Omega_c$ & $\Omega_b$\\\hline
$\left(\frac{1}{2}^+\right)$ & 3.062 & 6.397 & - & - & 3.234 & 6.511 \\\hline
\multirow{4}{*}{$\left(\frac{1}{2}^+,\frac{3}{2}^+\right)$} & (2.455, 2.519) & (5.833, 5.858) & (2.594, 2.649) & (5.970, 5.980) & (2.718, 2.776)& (6.081, 6.102)\\\cline{2-7}
& (2.958, 2.995) & (6.294, 6.326) & (3.136, 3.075) & (6.493, 6.376) & (3.152, 3.190) & (6.472, 6.478) \\\cline{2-7}
& (3.115, 3.116) & (6.447, 6.447) & - & - & (3.275, 3.280) & (6.593, 6.593)\\\cline{2-7}
& (3.182, 3.209) & - & - & - & (3.299, 3.321) & (6.648, 6.654) \\\hline
$\left(\frac{3}{2}^+,\frac{5}{2}^+\right)$ & (3.095, 3.108) & (6.426, 6.429) & - & - & (3.262, 3.273) & (6.576, 6.578) \\\hline
$\left(\frac{5}{2}^+,\frac{7}{2}^+\right)$ & (3.003, 3.015) & (6.325, 6.333) & - & - & - & (6.492, 6.497) \\\hline
$\left(\frac{1}{2}^-,\frac{3}{2}^-\right)$ & (2.848, 2.860) & (6.200, 6.202) & - & (6.305, 6.308) & (3.046, 3.056) & (6.388, 6.390) \\\hline
$\left(\frac{3}{2}^-,\frac{5}{2}^-\right)$ & (2.763, 2.790) & (6.101, 6.172) & (2.866, 2.895) & (6.190, 6.201) & (2.986, 3.014) & (6.304, 6.311) \\\hline

\end{tabular}
\end{table}
\end{center}

\section{Baryons with Two or Three Heavy Quarks}

\subsection{The $\Xi_{cc}$, $\Omega_{cc}$, $\Xi_{bb}$ and $\Omega_{bb}$ Baryons}

The Selex Collaboration has published an article in which the discovery of the $\Xi_{cc}$ with a mass of 3.519 GeV is reported \cite{selex}. Searches by the BaBar \cite{noselexa}, Belle \cite{noselexb} and Focus \cite{noselexc} Collaborations have all failed to confirm this state. We omit this state from Table \ref{multiplea} below, in which we show our results for the $\Xi_{cc}$, $\Omega_{cc}$, $\Xi_{bb}$ and $\Omega_{bb}$ baryons. We include this state Table \ref{baryonspec7}, which, along with Table \ref{baryonspec8}, shows the predictions of a number of authors for the masses of baryons containing two heavy quarks. 

If the candidate at 3.519 GeV is confirmed, describing such a state poses a challenge to models such as the one described herein, as most models give masses for the lowest lying $\Xi_{cc}$ that are in excess of 3.6 GeV. The notable exception is the model by Gerasyuta and co-authors \cite{Gerasyuta:1999pc,Gerasyuta:2007un,Gerasyuta:2008zy}, in which the states with negative parity are significantly lighter than those with positive parity, and the experimental candidate is assigned a $J^P=\fhlf^-$. We have tried to accommodate such a light $\Xi_{cc}$ in our model, but the resulting fit is significantly degraded in most other sectors. It is worth noting that models such as these usually are not this far wrong in predicting the masses of ground-state baryons in any sector, but the Coulomb interaction that we have in our model is vanishingly small. A value for $\alpha_{\rm Coul}$ as small as 0.1 (with no change in the other parameters) results in a mass for this state that is about 70 MeV lighter than the value shown in Table \ref{multiplea}, but this change leads to deterioration of the fit we have obtained in other sectors. We note that there has been one report of a $\Xi_{cc}$ state that is even lighter, with a mass of 3.460 GeV \cite{Moinester:2002uw}, as well as a heavier one with a mass of 3.78 GeV. If the lighter state is confirmed, most quark models, including the present work, will have to be modified to accommodate such a light state.

\begin{center}
\begin{table}[h]
\caption{Model predictions for $\Xi_{cc}$, $\Omega_{cc}$, $\Xi_{bb}$ and $\Omega_{bb}$. All masses are in GeV. The first column identifies the spin and parity of the state. 
\label{multiplea}}
\vspace{5mm}
\begin{tabular}{|l|c|c|c|c|}
\hline
$J^P$ & $\Xi_{cc}$ & $\Omega_{cc}$ & $\Xi_{bb}$ & $\Omega_{bb}$ \\\hline
\multirow{2}{*}{$\frac{1}{2}^+$} & 3.676 & 3.815 & 10.340 & 10.454 \\\cline{2-5}
& 4.029 & 4.180 & 10.576 & 10.693 \\\hline
\multirow{2}{*}{$\frac{3}{2}^+$} & 3.753 & 3.876 & 10.367 & 10.486 \\ \cline{2-5}
& 4.042 & 4.188 & 10.578 & 10.721 \\\hline
\multirow{2}{*}{$\frac{5}{2}^+$} & 4.047 & 4.202 & 10.676 & 10.720 \\ \cline{2-5}
& 4.091 & 4.232 & 10.712 & 10.734 \\\hline
\multirow{2}{*}{$\frac{7}{2}^+$} & 4.097 & 4.230 & 10.608 & 10.732 \\ \cline{2-5}
& 4.394 & 4.395 & 11.057 & 11.042 \\\hline
\multirow{2}{*}{$\frac{1}{2}^-$} & 3.910 & 4.046 & 10.493 & 10.616 \\ \cline{2-5}
& 4.074 & 4.135 & 10.710 & 10.763 \\\hline
\multirow{2}{*}{$\frac{3}{2}^-$} & 3.921 & 4.052 & 10.495 & 10.619 \\ \cline{2-5}
& 4.078 & 4.140 & 10.713 & 10.765 \\\hline
$\frac{5}{2}^-$ & 4.092 & 4.152 & 10.713 & 10.766\\\hline
\end{tabular}
\end{table}
\end{center}
\begin{center}
\begin{table}[h]
\caption{Model predictions for $\Xi_{cc}$, $\Xi_{cb}$ and $\Xi_{bb}$ in a number of quark models.
\label{baryonspec7}}
\vspace{5mm}
\begin{tabular}{|l|c|c|c|c|c|c|c|c|c|c|c|c|c|c|c|}\hline
State& Expt.& This&\cite{Albertus:2006ya}&\cite{SilvestreBrac:1996bg}&
\cite{Ebert:2002ig}&\cite{gershtein00,kiselev02}&\cite{narodetskii02}&\cite{Tong:1999qs}&\cite{itoh00}&\cite{vijande04}&\cite{Ebert:1996ec}&\cite{Roncaglia:1995az,roncaglia}&\cite{faessler}&\cite {Migura:2006ep}& \cite{Gerasyuta:1999pc,Gerasyuta:2007un,Gerasyuta:2008zy}\\
&&work&&&&&&&&&&&&&\\\hline
$\Xi_{cc}\hlf^+$ &3.519  &3.676&3.612&3.607&3.620&3.480&3.690&3.740&3.646&3.524&3.660&3.660&3.610&3.642&3.527\footnotemark[1]\\\hline   
$\Xi^*_{cc}\thlf^+$ &&3.753 &3.706 & &3.727&3.610&&3.860&3.733&3.548&3.810&3.740&3.680&3.723&3.597\\ \hline  
$\Xi_{cc}\hlf^-$ &&3.910&&&&&&&&&&&&3.920&3.410\\\hline
$\Xi_{cc}\thlf^-$ &&3.921&&&&&&&&&&&&3.920&3.140\\\hline
$\Xi_{cc}\fhlf^-$ &&4.092&&&&&&&&&&&&&3.519\footnotemark[1]\\\hline
$\Xi_{bc}$   &&7.011 &6.919 &6.915 &6.933&6.820&6.960&7.010&&&6.950&6.965&&&6.789\\ \hline  
$\Xi'_{bc}$  &&7.047 &6.948 &&6.963&6.850&&7.070&&&7.000&7.065&&&6.818\\\hline   
$\Xi^*_{bc}$ &&7.074&6.986 &&6.980&6.900&&7.100&&&7.020&7.060 &&&6.863\\ \hline  
$\Xi_{bb}$   &&10.340&10.197&10.194&10.202&10.090&10.160&10.300&&&10.230&10.340&&&10.045\\ \hline  
$\Xi^*_{bb}$ &&10.367 &10.236&&10.237&10.130&&10.340&&&10.280&10.370&&&10.104\\  \hline 
\end{tabular}
\footnotetext[1]{For these authors, a $\fhlf^-$ state at 3.519 GeV is chosen to match the experimental candidate state at the same mass}
\end{table}
\end{center}

\begin{center}
\begin{table}[h!!!]
\caption{Model predictions for $\Omega_{cc}$, $\Omega_{cb}$ and $\Omega_{bb}$ in a number of quark models.
\label{baryonspec8}}
\vspace{5mm}
\begin{tabular}{|l|c|c|c|c|c|c|c|c|c|c|c|c|c|}\hline
& This work&\cite{Albertus:2006ya}&\cite{SilvestreBrac:1996bg}&
\cite{Ebert:2002ig}&\cite{gershtein00,kiselev02}&\cite{narodetskii02}&\cite{Tong:1999qs}&\cite{itoh00}&\cite{Ebert:1996ec}&\cite{Roncaglia:1995az,roncaglia}&\cite{faessler}
&\cite {Migura:2006ep}& \cite{Gerasyuta:1999pc,Gerasyuta:2007un,Gerasyuta:2008zy}\\\hline
$\Omega_{cc}$   &3.815 &3.702  &3.710&3.778&3.590&3.860&3.760&3.749&3.760&3.740&3.710&3.732&3.598\\\hline   
$\Omega^*_{cc}$ &3.876&3.783 &&  3.872&3.690&&3.900&3.826&3.890&3.820&3.760&3.765&3.700\\\hline   
$\Omega_{bc}$   &7.136 &6.986&7.003&7.088&6.910&7.130&7.050&&7.050&7.045&&&6.798\\\hline   
$\Omega'_{bc}$  &7.165 &7.009& &7.116&6.930&&7.110&&7.090&7.105&&&6.836\\\hline   
$\Omega^*_{bc}$ &7.187&7.046 &&7.130&6.990&&7.130&&7.110&7.120& &&6.914\\\hline   
$\Omega_{bb}$   &10.454&10.260 &10.267&10.359&10.180&10.340&10.340&&10.320&10.370&&&9.999\\ \hline  
$\Omega^*_{bb}$ &10.486 &10.297 &&10.389&10.200&&10.380&&10.360&10.400&&&10.126\\   \hline
\end{tabular}
\end{table}
\end{center}

\begin{center}
\begin{figure}[h]
\includegraphics[width=5.5in]{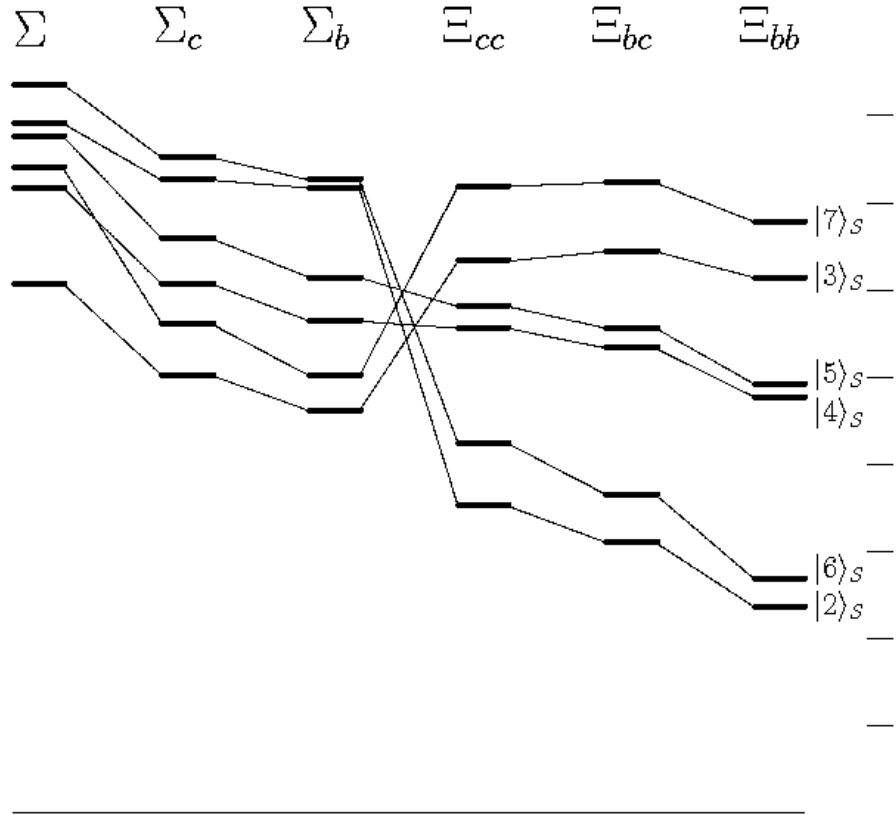}
%\vspace*{-1.0in}
\caption{Energy levels of the six excitations in the $\hlf^+$ spectrum for the $\Sigma$, $\Sigma_c$, $\Sigma_b$, $\Xi_{cc}$, $\Xi_{bc}=(cb+bc)u$ and $\Xi_{bb}$ states, respectively. All spectra are shown relative to the ground state in each sector, and the positions of the lines are drawn to scale. The lines on the right of the diagram indicate 100 MeV intervals, and the notation $|i\rangle_{\bf S}$ on the diagram indicates which of the wave function components of Table \ref{wfcomponents} dominates the wave function of that state.}
\label{energylevels}
\end{figure}
\end{center}

One feature of our results not apparent from the results in this table is the hierarchy that occurs in the excited states. Excitations in these states can arise from an excitation in the $\rho$ coordinate (in the `diquark' made up of the two heavy quarks), or in the $\lambda$ coordinate (corresponding to an excitation in the relative coordinate between the heavy diquark and the light quark). If we examine the excitations for the set of states with a particular $J^P$, say $\hlf^+$, we find that there is a clear ordering of the excitations depending on whether the excitation is in $\rho$ or $\lambda$. This hierarchy is most easily discussed in terms of an energy-level diagram, shown in Fig. \ref{energylevels}.

For these $\hlf^+$ states, the seven components of the wave function are shown in the last column of Table \ref{wfcomponents}. The second and sixth of these components have excitations in $\rho$, the third and seventh have excitations in $\lambda$, and the fourth and fifth have excitations in both $\rho$ and $\lambda$. In Fig. \ref{energylevels}, we show the energies of the six excitations in the $\hlf^+$ spectrum for the $\Sigma$, $\Sigma_c$, $\Sigma_b$, $\Xi_{cc}$, $\Xi_{bc}=(cb+bc)u$ and $\Xi_{bb}$ states, respectively. All spectra are shown relative to the ground state in each sector, and the positions of the lines are drawn to scale. The lines on the right of the diagram indicate 100 MeV intervals, and the notation $|i\rangle_{\bf S}$ on the diagram indicates  which of the wave function components of Table \ref{wfcomponents} dominates the wave function of that model state (the ${\bf S}$ indicates that wave function components are those for a baryon with flavor wave function symmetric in the first two quarks, or the sextets in Table \ref{wfcomponents}).

When the `diquark' in the baryon is composed of light quarks, with the heavy quark as the third quark, excitations in the $\lambda$ coordinate cost less energy than those in the $\rho$ coordinate, leading to the ordering in the spectrum seen in the $\Sigma_c$ and $\Sigma_b$. Here, we include the $\Sigma$ to illustrate that, when the baryon consists of only light quarks, excitations in either coordinate cost similar amounts of energy, leading to a spectrum in which the excited states are very close together, and there is no obvious ordering of $\rho$ and $\lambda$ excitations. When the diquark is heavy, the $\rho$ coordinate costs less energy to excite, and the ordering of states becomes `inverted', with the $\lambda$ excitations becoming heavier than the $\rho$ excitations. This contradicts many treatments of these states that assume that the heavy diquark is `tightly bound' and difficult to excite. It is also interesting to note that, as the diquark gets heavier, the energy differences between states with different kinds of excitations becomes larger. This hierarchy of states is repeated for all values of $J^P$ that we have examined in the model. For the $\Lambda_Q$, this hierarchy of excitations is not as easy to identify as it is with other states, in large part due to the large role played by the contact hyperfine interaction.

\subsection{The $\Omega_{bcc}$, $\Omega_{ccc}$, $\Omega_{bbc}$ and $\Omega_{bbb}$ Baryons}

Our predictions for the spectra of $\Omega_{bcc}$, $\Omega_{ccc}$, $\Omega_{bbc}$ and $\Omega_{bbb}$ baryons are shown in Table \ref{multipleb}. For the $\Omega_{ccc}$ and $\Omega_{bbb}$, the symmetry of the flavor wave function requires the spin-space wave function to be fully symmetric, leading to a different counting of states in the spectrum. This is reflected in the blank lines in the table. Nevertheless, the table does not show that there are far fewer of these states (up to the $N=2$ oscillator band) than there are $\Omega_{bcc}$ states, say. For $J^P=\hlf^+$, there are seven of the latter states, but only two $\Omega_{QQQ}$ states. 

Our predicted masses for the lowest lying states in each of these sectors are somewhat heavier than in other calculations \cite{bjorken,faessler,martynenko}. One possible source for this difference may be our essentially non-existent Coulomb interaction. These heavy quarks are expected to reside close to each other, when the Coulomb interaction would make a significant contribution to their `binding'. In our model, the fits to the spectrum lead to a negligibly small Coulomb interaction, and this gives rise to some ground states that are too heavy. This effect would be expected to be smaller in the excited states, as the average separation of the quarks is increased.

\begin{center}
\begin{table}[h]
\caption{Model predictions for $\Omega_{bcc}$, $\Omega_{ccc}$, $\Omega_{bbc}$, and $\Omega_{bbb}$ baryons. All masses are in GeV. The first column identifies the spin and parity of the state. 
\label{multipleb}}
\vspace{5mm}
\begin{tabular}{|l|c|c|c|c|}
\hline
$J^P$ & $\Omega_{bcc}$ & $\Omega_{ccc}$ & $\Omega_{bbc}$ &  $\Omega_{bbb}$ \\\hline
\multirow{2}{*}{$\frac{1}{2}^+$} & 8.245 & 5.325 & 11.535 & 15.097 \\ \cline{2-5}
& 8.537 & 5.332 & 11.787 & 15.102 \\\hline
\multirow{2}{*}{$\frac{3}{2}^+$} & 8.265 & 4.965 & 11.554 & 14.834 \\ \cline{2-5}
& 8.553 & 5.313 & 11.798 & 15.089 \\\hline
\multirow{2}{*}{$\frac{5}{2}^+$} & 8.568 & 5.329& 11.823 &15.101 \\ \cline{2-5}
& 8.571 & 5.343 & 11.831 & 15.109 \\\hline
\multirow{2}{*}{$\frac{7}{2}^+$} & 8.568 & 5.331 & 11.810 & 15.101 \\\cline{2-5}
& 8.653 & - & 11.908 & - \\\hline
\multirow{2}{*}{$\frac{1}{2}^-$} & 8.418 & 5.155& 11.710 & 14.975 \\ \cline{2-5}
& 8.422 & - &11.757 & - \\\hline
\multirow{2}{*}{$\frac{3}{2}^-$} & 8.420 & 5.160& 11.711 & 14.976 \\ \cline{2-5}
& 8.422 & - & 11.759 & - \\\hline
$\frac{5}{2}^-$ & 8.432 & -& 11.762 & - \\ \hline
\end{tabular}
\end{table}
\end{center}

\subsection{The $\Xi_{bc}$ and $\Omega_{bc}$ Baryons}

The $\Xi_{bc}$ and $\Omega_{bc}$ baryons belong in a triplet of SU(3) comprising the $\Xi_{bc}^0\,\,(bcd)$, the $\Xi_{bc}^+\,\,(bcu)$ and the $\Omega_{bc}^0\,\,(bcs)$. As mentioned in an earlier section, it has been argued that for these states, the pair of heavy quarks bind into a pointlike diquark that can have spin zero or one, with the two possible diquark spins being conserved in many treatments. In our model, we treat these states in this way, and present spectra that are obtained when no mixing is allowed, as well as when mixing is turned on. Since we treat a number of excited states, it is not accurate for us to refer to our states as being built from scalar and axial-vector diquarks. Instead, we'll refer to them as members of the `triplet' and `singlet' of a (pseudo)symmetry group, SU(2)$_{bc}$, where the subscript denotes the quark flavors that are used to construct the singlet and triplet representations.

This SU(2)$_{bc}$ symmetry is broken, and the $\Xi_{bc}$ states of the singlet and triplet should be mixed, unless the operators responsible for such mixing are suppressed by the masses of the heavy quarks. However, as noted when we discussed the mixing in the $\Xi_c$ and $\Xi_b$ systems, some of the mixing arises from the linear and Coulomb terms in the Hamiltonian, and do not vanish when the quark masses get large. In addition, even though the hyperfine contributions to mixing get small, so do their contributions to the diagonal matrix elements, and it is the relative sizes of the diagonal and mixing terms that ultimately determine the mixing angles. Thus, there is no reason to expect small mixing angles in this sector. 

Our results for the $\Xi_{bc}$ states are shown in Tables \ref{cascadebc}. The spin and parity of the states are shown in column one. Column two shows the masses that result when the states are treated as being purely flavor singlet or triplet in SU(2)$_{bc}$, while column three shows the masses when mixing between these two representations is allowed. Column four shows the dominant contributions to the wave functions when there's no mixing, while column five shows this when mixing is allowed. Column six shows a singlet-triplet mixing angle as defined in Eq. (\ref{angle}). Note that the spin-space wave functions for the SU(2)$_{bc}$ singlet are the same as those for the SU(3)$_{uds}$ antitriplet, and the spin-space wave functions for the SU(2)$_{bc}$ triplet are the same as those for the SU(3)$_{uds}$, sextet. Both sets of spin-space wave functions are shown in Table \ref{wfcomponents}.

\begin{center}
\begin{table}[h]
\caption{The $\Xi_{bc}$ spectrum obtained in our model. The first column identifies the state (the spin and parity are the results of the model). The second column shows the masses when no mixing is allowed between the {\bf 3} and ${\bf 1}$ representations of SU(2)$_{bc}$, while the third column results when such mixing is allowed. The fourth  column shows the dominant components in the wave function when there is no mixing, while the fifth column shows the same when mixing is allowed. The last column is an estimate of the mixing angle between the two SU(2) representations, using a definition analogous to that given in Eq. (\ref{angle}). All masses are in GeV.
\label{cascadebc}}
\vspace{5mm}
\begin{tabular}{|c|c|c|c|c|c|}
\hline
$J^P$& \multicolumn{2}{c|}{Model}&\multicolumn{2}{c|}{Dominant Wave Function Components}& $\left|\tan\phi\right|$ \\\cline{2-5}
& Unmixed & Mixed & Unmixed & Mixed & (Eq. \ref{angle})\\\hline
\multirow{4}{*}{$\frac{1}{2}^+$}&7.020&7.011& $0.9881 |1\rangle_{\bf 3}+0.1486 |2\rangle_{\bf 3}$
&$0.8976 |1\rangle_{\bf 3}+0.3839|1\rangle_{\bf 1}$&0.4375\\
 &&&
$+0.0314 |4\rangle_{\bf 3}+0.0242 |3\rangle_{\bf 3}$&
$+0.1526|2\rangle_{\bf 3}+0.1047 |4\rangle_{\bf 1}$&\\\cline{2-6}

& 7.044& 7.047 & $0.9885 |1\rangle_{\bf 1}+0.1156|2\rangle_{\bf 1}$ & $0.9013 |1\rangle_{\bf 1}-0.3908 |1\rangle_{\bf 3}$ & 0.4398 \\ 
&&& $+0.0937|3\rangle_{\bf 1}+0.0258 |4\rangle_{\bf 1}$ & $+0.1636 |2\rangle_{\bf 1}+0.0649 |4\rangle_{\bf 3}$ & \\\hline

\multirow{6}{*}{$\frac{3}{2}^+$} & 7.078&7.074 & $0.9602|1\rangle_{\bf 3} +0.2789|2\rangle_{\bf 3} $ &$0.9633|1\rangle_{\bf 3}+0.2264 |2\rangle_{\bf 3} $& 0.0721\\ 
 & & & $-0.0102|3\rangle_{\bf 3}$ &$-0.1249|3\rangle_{\bf 3}+0.0719|1\rangle_{\bf 1} $&\\\cline{2-6}

& 7.386& 7.371& $0.9974|6\rangle_{\bf 3} -0.0462|7\rangle_{\bf 3} $ &$0.9643|6\rangle_{\bf 3}-0.1990 |6\rangle_{\bf 1} $&0.2607\\ 
& & & $-0.0392|5\rangle_{\bf 3}$ &$+0.1379|4\rangle_{\bf 1}-0.0812 |7\rangle_{\bf 3} $&\\\cline{2-6}

& 7.369 & 7.397 & $0.9984|4\rangle_{\bf 1} -0.0544|6\rangle_{\bf 1} $ &$0.9661|4\rangle_{\bf 1}-0.1612 |7\rangle_{\bf 3} $&0.2506\\ 
 & &&  $+0.0136|7\rangle_{\bf 1}$ &$-0.1548|6\rangle_{\bf 3}+0.1030 |2\rangle_{\bf 3} $&\\\hline

\multirow{4}{*}{$\frac{5}{2}^+$} & 7.356 &7.368 & $0.9990|3\rangle_{\bf 3}-0.0368 |1\rangle_{\bf 3} $ &$0.9683|3\rangle_{\bf 3}-0.1883 |3\rangle_{\bf 1} $&0.2376\\ 
& & & $+0.0186|2\rangle_{\bf 3} +0.0180|5\rangle_{\bf 3} $ &$+0.1326|4\rangle_{\bf 1}-0.0806|1\rangle_{\bf 3} $&\\\cline{2-6}

& 7.374 &7.396 & $0.9991|4\rangle_{\bf 1}-0.0415 |3\rangle_{\bf 1} $ &$0.9744|4\rangle_{\bf 1} -0.1538|3\rangle_{\bf 3} $&0.2249\\ 
& && $+0.0120|5\rangle_{\bf 1}$ &$-0.1492|1\rangle_{\bf 3}-0.0457 |2\rangle_{\bf 3} $& \\\hline

\multirow{4}{*}{$\frac{7}{2}^+$} &7.415 &7.375 & $|1\rangle_{\bf 3} $ &$0.9845|1\rangle_{\bf 3} -0.1736|1\rangle_{\bf 1} $&0.1763\\ 
& & & &$-0.0233|2\rangle_{\bf 3}$&\\\cline{2-6}

&7.564 &7.562 & $|1\rangle_{\bf 1}$ &$0.9637|1\rangle_{\bf 1}-0.2022 |2\rangle_{\bf 3} $&0.2074\\ 
& & & &$+0.1747|1\rangle_{\bf 3}$&\\\hline

\multirow{4}{*}{$\frac{1}{2}^-$} & 7.206& 7.227& $0.9964|2\rangle_{\bf 1}-0.0785 |1\rangle_{\bf 1} $ &$0.9723|2\rangle_{\bf 1}+0.1533 |1\rangle_{\bf 3} $&0.2177\\ 
& & & $-0.0327|3\rangle_{\bf 1}$ &$-0.1473|3\rangle_{\bf 3}-0.0772 |3\rangle_{\bf 1} $&\\\cline{2-6}

& 7.231& 7.267& $0.9994|1\rangle_{\bf 3} -0.0332|3\rangle_{\bf 3} $ &$0.9750|1\rangle_{\bf 3}-0.1633 |2\rangle_{\bf 1} $&0.2224\\ 
& & & &$-0.1054|3\rangle_{\bf 1}+0.0968 |1\rangle_{\bf 1} $&\\\hline

\multirow{4}{*}{$\frac{3}{2}^-$} & 7.208&7.217 &$0.9991|2\rangle_{\bf 1}-0.0329 |3\rangle_{\bf 1} $ &$0.9665|2\rangle_{\bf 1}+0.1846 |1\rangle_{\bf 3} $&0.2514 \\ 
& & & $+0.0281|1\rangle_{\bf 1}$ &$-0.1593|3\rangle_{\bf 3}-0.0780 |3\rangle_{\bf 1} $&\\\cline{2-6}

& 7.229& 7.252& $0.9994|1\rangle_{\bf 3} -0.0354|3\rangle_{\bf 3} $ &$0.9706|1\rangle_{\bf 3}-0.1983 |2\rangle_{\bf 1} $&0.2459\\ 
& & & $-0.0108|2\rangle_{\bf 3}$ &$-0.1192|3\rangle_{\bf 1} -0.0573|1\rangle_{\bf 1} $&\\\hline

\multirow{2}{*}{$\frac{5}{2}^-$} & 7.272&7.290 & $|1\rangle_{\bf 1}$ &$0.9951|1\rangle_{\bf 1}-0.0990 |1\rangle_{\bf 3} $&0.0995\\[+5pt]\cline{2-6}

& 7.414&7.509 & $|1\rangle_{\bf 3}$ &$0.9951|1\rangle_{\bf 3}+0.0990 |1\rangle_{\bf 1}$&0.0995\\[+5pt]\hline
\end{tabular}
\end{table}
\end{center}

One of the first things to note about the results in Table \ref{cascadebc} is the fact that the ordering of states differs from the ordering when the baryon contains two light quarks. Among the $\Xi_b$ and $\Xi_c$, the lowest lying state in all treatments that we know of is the one that belongs to the antitriplet (antisymmetric in the $q$ and $s$ quarks), while in Table \ref{cascadebc}, the lowest lying state belongs to the SU(2)$_{bc}$ analog of the sextet (symmetric in the $b$ and $c$ quarks). This `inversion' seems to occur in all treatments of these states.

The second point to note about this table is the very large `mixing angles' between triplet and singlet components that occur for most states. As we have noted, this should not be too surprising, as the approximate flavor symmetry assumed is not very close to being realized in nature. Note, too, that while mixing may change the masses of states by only a few tens of MeV at most, the effects on their wave functions is much more significant. Properties such as the electroweak and strong decays of these states can be expected to show significant deviations from those predicted using the SU(2) symmetry. One state provides a notable exception to this general trend: the mixing angle for the lowest lying $J^P=\thlf^+$ is quite small.

One of the more intriguing results in the table occurs in the $J^P=\thlf^+$ sector. There, when mixing is turned off, the lowest lying state belongs to the triplet, while the next lowest lying state lies in the singlet. When mixing is turned on, this second state is pushed higher in mass, and becomes the third lowest lying state, while another (predominantly) triplet state usurps its position as the second lowest state. Thus, mixing not only affects the masses of the states but also their ordering, in some cases.

The $\Omega_{bc}$ spectrum is shown in Table \ref{omegabc}. Much of what was noted for the $\Xi_{bc}$ can be repeated here: mixing angles are generally large except for the case of the lowest lying $\thlf^+$ state, the lowest lying state belongs to the triplet, not the singlet as the spectrum of light hadrons might lead us to expect, and other orderings of states are significantly changed. In this sector, the lowest lying, predominantly singlet state with $J^P=\thlf^+$ is the fourth one, both with and without mixing. This is already quite different from the $\Xi_{bc}$ case, and shows a significant departure from the sector with a single heavy quark in the baryon.

\begin{center}
\begin{table}
\caption{The $\Omega_{bc}$ spectrum obtained in our model. The key to the columns is as in Table \ref{cascadebc}. All masses are in GeV.
\label{omegabc}}
\begin{tabular}{|c|c|c|c|c|c|}\hline
%\vspace{5mm}
%\begin{tabular}{|c|c|c|c|c|c|c|}
%
$J^P$& \multicolumn{2}{c|}{Model}&\multicolumn{2}{c|}{Dominant Wave Function Components}& $\left|\tan\phi\right|$ \\\cline{2-5}
& Unmixed & Mixed & Unmixed & Mixed &(Eq. \ref{angle}) \\\hline
\multirow{4}{*}{$\frac{1}{2}^+$}&7.147&7.136& $0.9858 |1\rangle_{\bf 3}+0.1593 |2\rangle_{\bf 3}$
&$0.8947 |1\rangle_{\bf 3}+0.4149|1\rangle_{\bf 1}$&0.4789\\
 &&&
$+0.0429 |3\rangle_{\bf 3}+0.0323 |4\rangle_{\bf 3}$&
$+0.1124|4\rangle_{\bf 1}+0.0868 |3\rangle_{\bf 3}$&\\\cline{2-6}
& 7.166& 7.165 & $0.9793 |1\rangle_{\bf 1}+0.1877|2\rangle_{\bf 1}$ & $0.9020 |1\rangle_{\bf 1}-0.4205 |1\rangle_{\bf 3}$ & 0.4722 \\ 
 &&& $+0.0695|3\rangle_{\bf 1}+0.0286 |4\rangle_{\bf 1}$ & $+0.0674 |4\rangle_{\bf 3}+0.0517 |2\rangle_{\bf 1}$ & \\\hline

\multirow{4}{*}{$\frac{3}{2}^+$} & 7.191&7.187 & $0.9921|1\rangle_{\bf 3}+0.1111 |3\rangle_{\bf 3} $ &$0.9923|1\rangle_{\bf 3}+0.0849 |1\rangle_{\bf 1} $&0.0853 \\ 
 & & & $+0.0581|2\rangle_{\bf 3}$ &+0.0801$|2\rangle_{\bf 3}-0.0392 |3\rangle_{\bf 3} $&\\\cline{2-6}

& 7.487&7.467 & $0.9918|6\rangle_{\bf 3}-0.1070 |2\rangle_{\bf 3} $ &$0.9220|6\rangle_{\bf 3}-0.2772 |6\rangle_{\bf 1} $&0.3813\\ 
 & &&  $-0.0542|7\rangle_{\bf 3}+0.0327 |9\rangle_{\bf 3} $ &$+0.2207|4\rangle_{\bf 1} -0.1230|7\rangle_{\bf 3} $&\\\hline

\multirow{4}{*}{$\frac{5}{2}^+$} & 7.479&7.467 & $0.9969|3\rangle_{\bf 3}-0.0679 |1\rangle_{\bf 3} $ &$0.9250|3\rangle_{\bf 3}-0.2727 |3\rangle_{\bf 1} $&0.3762\\ 
& & & $+0.0358|5\rangle_{\bf 3} +0.0200|2\rangle_{\bf 3} $ &$+0.2201|4\rangle_{\bf 1}-0.1209 |1\rangle_{\bf 3} $&\\\cline{2-6}

& 7.498& 7.490& $0.9963|4\rangle_{\bf 1}-0.0833 |3\rangle_{\bf 1} $ &$0.9378|4\rangle_{\bf 1}-0.2557 |3\rangle_{\bf 3} $&0.3656\\ 
& && $+0.0217|5\rangle_{\bf 1}$ &$-0.2208|1\rangle_{\bf 3}-0.0612 |2\rangle_{\bf 3} $& \\\hline

\multirow{4}{*}{$\frac{7}{2}^+$} & 7.509 & 7.498 & $0.9999|1\rangle_{\bf 3}+0.0125 |2\rangle_{\bf 3} $ &$0.9688|1\rangle_{\bf 3}-0.2436 |1\rangle_{\bf 1} $&0.2511\\ 
& & & &$+0.0467|2\rangle_{\bf 3}$&\\\cline{2-6}

&7.633 &7.619 & $|1\rangle_{\bf 1}$ &$0.9197|1\rangle_{\bf 1}-0.3062 |2\rangle_{\bf 3} $&0.4270\\ 
& & & &$+0.2460|1\rangle_{\bf 3} |3\rangle_{\bf 3} $&\\\hline

\multirow{4}{*}{$\frac{1}{2}^-$} & 7.335 & 7.320 & $0.9937|2\rangle_{\bf 1} -0.0944|1\rangle_{\bf 1} $ &$0.9320|2\rangle_{\bf 1}+0.2465 |1\rangle_{\bf 3} $&0.3619\\ 
& & & $-0.0603|3\rangle_{\bf 1}$ &$-0.2343|3\rangle_{\bf 3}-0.1095 |3\rangle_{\bf 1} $&\\\cline{2-6}

 &7.346 &7.343 & $0.9960|1\rangle_{\bf 3}-0.0891 |3\rangle_{\bf 3} $ &$0.9302|1\rangle_{\bf 3}-0.2594 |2\rangle_{\bf 1} $&0.3929\\ 
& & &  &$-0.2086|3\rangle_{\bf 1}+0.1515 |1\rangle_{\bf 1} $&\\\hline

\multirow{4}{*}{$\frac{3}{2}^-$} &7.334 &7.322& $0.9976|2\rangle_{\bf 1}-0.0616 |3\rangle_{\bf 1} $ &$0.9324|2\rangle_{\bf 1}+0.2503 |1\rangle_{\bf 3} $&0.3657\\ 
& & & $+0.0328|1\rangle_{\bf 1}$ &$-0.2352|3\rangle_{\bf 3}-0.1107 |3\rangle_{\bf 1} $&\\\cline{2-6}

& 7.349&7.345 & $0.9960|1\rangle_{\bf 3}-0.0894 |3\rangle_{\bf 3} $ &$0.9353|1\rangle_{\bf 3}-0.2723 |2\rangle_{\bf 1} $&0.3778\\ 
& & &  &$-0.2095|3\rangle_{\bf 1}-0.0829 |1\rangle_{\bf 1} $&\\\hline

\multirow{2}{*}{$\frac{5}{2}^-$} &7.362 &7.356 & $|1\rangle_{\bf 1}$ &$0.9772|1\rangle_{\bf 1}-0.2134 |1\rangle_{\bf 3} $&0.2173\\ \cline{2-6}

 &7.517 &7.468 & $|1\rangle_{\bf 3}$ &$0.9772|1\rangle_{\bf 3}+0.2134 |1\rangle_{\bf 1} $&0.2173\\ \hline
\end{tabular}
\end{table}
\end{center}

In the two lowest lying states of both the $\Xi_{bc}$ and the $\Omega_{bc}$ spectra, the mixing between the two dominant components of the wave function is highly suggestive of the result when a $(2\leftrightarrow 3)$ permutation is carried out on a spin wave function of type $\chi_\rho$ or $\chi_\lambda$. The exact results of such a permutation are
\beq
\{23\} \chi^\rho(s)=\frac{\sqrt{3}}{2}\chi^{\lambda}(s)+\frac{1}{2}
\chi^{\rho}(s),\,\,\,\,
\{23\} \chi^\lambda(s)=\frac{\sqrt{3}}{2}\chi^{\rho}(s)-\frac{1}{2}
\chi^{\lambda}(s),
\eeq
and $\sqrt{3}/2=0.8660$, close to the coefficient of the $|1\rangle_{\bf 3}$ component of the lowest lying state, as well as to the coefficient of the $|1\rangle_{\bf 1}$ of the second lowest state with $J^P=\hlf^+$, for both $\Xi_{bc}$ and $\Omega_{bc}$. These two components are both symmetric in space (they both correspond to $l_\rho=l_\lambda=n_\rho=n_\lambda=0$), with $|1\rangle_{\bf 3}\simeq \chi_\lambda$ and $|1\rangle_{\bf 1}\simeq\chi_\rho$.

If we treat these baryons in a basis where the flavor wave functions are $\Xi_{uc}^{\bf 1}=1/\sqrt{2}(uc-cu)b$ and $\Xi_{uc}^{\bf 3}=1/\sqrt{2}(uc+cu)b$, and allow these representations to mix, the wave functions and mixing angles that result are shown in Table \ref{newmix}. In that table, we refer to the flavor wave function $1/\sqrt{2}(uc-cu)b$ as the singlet representation, and $1/\sqrt{2}(uc+cu)b$ as the triplet. 
\begin{center}
\begin{table}
\caption{Selected $\Xi_{bc}$ and $\Omega_{bc}$ wave functions and mixing angles that result when the states are treated as $1/\sqrt{2}(qc\pm cq)b$, where $q$ denotes a light quark.
\label{newmix}}
\begin{tabular}{|c|c|c|c|c|c|}\hline
\multicolumn{3}{|c|}{$\Xi_{bc}$}&\multicolumn{3}{c|}{$\Omega_{bc}$}\\\hline
State & Dominant Wave & $\left|\tan\phi\right|$& State & Dominant Wave & $\left|\tan\phi\right|$\\
& Function Components & (Eq. \ref{angle})& & Function Components& (Eq. \ref{angle})\\\hline
$\Xi_{bc}\hlf^+(7.011)$& $0.9888|1\rangle_{\bf 1}+0.1051 |4\rangle_{\bf 3} $& 0.1382 & $\Omega_{bc}\hlf^+(7.136)$& $0.9727|1\rangle_{\bf 1}+0.1490 |2\rangle_{\bf 1} $& 0.1031 \\
& & & & $+0.1438|3\rangle_{\bf 1}$ &\\\hline
$\Xi_{bc}\hlf^+(7.047)$& $0.9790|1\rangle_{\bf 3}+0.1249 |4\rangle_{\bf 1} $& 0.1573 & $\Omega_{bc}\hlf^+(7.165)$& $0.9763|1\rangle_{\bf 3}+0.1673 |3\rangle_{\bf 3} $& 0.1153 \\
& $-0.1086|2\rangle_{\bf 3}$ &&&&\\\hline

$\Xi_{bc}\thlf^+(7.074)$& $0.9374|1\rangle_{\bf 3}+0.3090 |2\rangle_{\bf 3} $& 0.0942 & $\Omega_{bc}\thlf^+(7.187)$& $0.9874|1\rangle_{\bf 3}+0.1070 |2\rangle_{\bf 3} $& 0.0944 \\
&$+0.1307|3\rangle_{\bf 3}$ &&&&\\\hline
$\Xi_{bc}\thlf^+(7.397)$&$0.9301|7\rangle_{\bf 1}-0.2731 |7\rangle_{\bf 3} $ & 0.3797& $\Omega_{bc}\thlf^+(7.467)$& $0.9538|7\rangle_{\bf 1}-0.2642 |7\rangle_{\bf 3} $& 0.3017  \\
& $+0.1664|3\rangle_{\bf 3}$& & & $+0.1133|9\rangle_{\bf 3}$ &\\\hline
\end{tabular}
\end{table}
\end{center}

The results in Table \ref{newmix} show that the two lowest $J^P=\hlf^+$ states are better described as states in the single and triplet representations of SU(2)$_{qc}$, where $q$ is an up, down or strange quark, than they are as states in the corresponding representations of SU(2)$_{bc}$, {\em particularly when mixing between the two representations is ignored}. This conclusion is in agreement with that of Franklin {\it et al.} \cite{franklin}, who pointed out that the best choice of quark orderings is to pick quarks 1 and 2 to give the smallest mass difference $m_2-m_1$, as this leads to the smallest mixing angle. Obviously, if mixing is allowed, one choice of basis is as good as another. Among the higher excited states, the choice of either basis leads to moderately large mixing angles. For the $\thlf^+$ states, it appears that either basis works well for the ground state, but the next lowest state slightly prefers the $1/\sqrt{2}(qc\pm cq)b$. In this basis, the lowest lying state has flavor wave function $1/\sqrt{2}(qc- cq)b$ (and, the largest component has total spin in the $qc$ diquark as zero), more in keeping with the baryons with a single heavy quark. Thus, the fact that the lowest lying state in the $1/\sqrt{2}(bc\pm cb)q$ basis has spin one in the $bc$ diquark is an artifact that arises because mixing isn't allowed between the representations in the basis. It is worth noting that the singlet component of the lowest-lying state shown in Table \ref{newmix} is not unusual for a state with wave function that is antisymmetric in the first two quarks. For instance, the dominant components of the $\Lambda_c(2285)\hlf^+$ are $0.9750 |1\rangle_{\bf \overline{3}}+0.1943 |2\rangle_{\bf \overline{3}}+0.1026 |3\rangle_{\bf \overline{3}}$. Similarly, for the $\Sigma_c(2455)\hlf^+$, the dominant wave function components are $0.9868 |1\rangle_{\bf 6}+0.1141 |2\rangle_{\bf 6}+0.1062 |3\rangle_{\bf \overline{3}}$, while for the $\Sigma_c(2519)\thlf^+$, they are  $0.9983 |1\rangle_{\bf 6}+0.0571 |3\rangle_{\bf 6}$. 

It is useful to try to understand the mixing that arises in a simplified version of our model. Let us consider the two lowest-lying states in the $J^P=\hlf^+$ sector, and treat them in terms of single-component spin-space wave functions. Defining $|\rho\rangle=\chi_\rho$ and $|\lambda\rangle=\chi_\lambda$, and the two lowest-lying eigenstate as
\beqy 
|1\rangle&=&\cos{\phi} |\lambda\rangle+\sin{\phi} |\rho\rangle,\nonumber\\
|2\rangle&=&\cos{\phi} |\rho\rangle-\sin{\phi} |\lambda\rangle,
\eeqy
the Hamiltonian matrix that arises is 
\beq
H_{\rho\lambda}=\left(\begin{tabular}{cc}
${\cal H}_0+\Delta_1$&$-\delta_m$\\
$-\delta_m$ & $H_0-\Delta_2$\end{tabular}\right).
\eeq
Here ${\cal H}_0$ is the leading order mass of the originally degenerate pair of states, $\Delta\equiv\Delta_2+\Delta_1$ is the mass splitting that results from the hyperfine interaction, and $\delta_m$ is the matrix element of the hyperfine interaction that mixes the two states. Diagonalizing this matrix leads to a mass splitting $\Delta_m$ between the states that is
\beq
\Delta_m=\sqrt{\Delta^2 + 4\delta_m^2},
\eeq
and a mixing angle given by
\beq
\tan{\phi}=\frac{-\Delta + \sqrt{\Delta^2 + 4\delta_m^2}}{2\delta_m}.
\eeq
In our model, $\Delta$ is proportional to $1/(m_b m_c)$, while $\delta_m$ has two contributions, one proportional to $1/(m_b m_u)$, the second proportional to $1/(m_c m_u)$, with a relative negative sign between them. In any case, the fact that $\Delta$ is expected to be smaller than $\delta_m$, perhaps significantly so, means that $\tan{\phi} \lesssim 1$, indicating potentially large mixing angles. If we explicitly use 
\beqy\label{mixeqn}
\Delta_1&=&\frac{\gamma_{12}}{4 m_1 m_2}-\frac{1}{2 m_3}\left(\frac{\gamma_{13}}{m_1}+\frac{\gamma_{23}}{m_2}\right),\nonumber\\
\Delta_2&=&\frac{3 \gamma_{12}}{4 m_1 m_2},\nonumber\\
\delta_m&=&-\frac{\sqrt{3}}{4 m_3}\left(\frac{\gamma_{13}}{m_1}-\frac{\gamma_{23}}{m_2}\right),
\eeqy
as would be the case in the quark model, Eq. (\ref{hyp}), with $\gamma_{ij}$ being the values of the matrix element of the spatial parts of the operator, the mixing angle that results is given by
\beq
\tan{\phi}=\frac{m_1 + m_2 - 2m_3 - 2\sqrt{m_1^2 + m_2^2 + m_3^2-m_1m_2-m_1 m_3 - m_2m_3}}{\sqrt{3}(m_1 - m_2)},
\eeq
in the limit that $\gamma_{12}=\gamma_{13}=\gamma_{23}$. Using the masses that result from our fit to the spectrum, and putting the up quark as the third quark, this yields a value of $\tan{\phi}\approx 0.386$, somewhat smaller than but similar to the mixing angle shown in the first row of Table \ref{cascadebc}. If, on the other hand, the $b$ quark is selected as the third quark, the value of $\tan{\phi}$ obtained is 0.156, slightly larger than the mixing angle shown in the first row of Table \ref{newmix}. The mixing angles calculated in this simple model indicate that the values obtained in the full model are feasible.

Our results for the $\Xi_{bc}$ and $\Omega_{bc}$ imply that, in the heavy quark expansion, treating these states and their decays at leading order could lead to misleading results. The $1/m_c$ and $1/m_b$ corrections to the masses of the states are small, but the corrections to the wave functions are not necessarily so. In the heavy quark expansion, the first diagonal corrections to the masses of the states ($\Delta_{1,2}$ in Eq. (\ref{mixeqn})) from the hyperfine interaction appears at order $1/(m_b m_c)$, while the off-diagonal term, the one that leads to mixing, appears at order $1/m_c$ or $1/m_b$. In the expression for $\tan{\phi}$, in the limit that the two heavy quark masses get very large, $\tan{\phi}$ can approach unity.

The mixing in the $\Xi_{bc}$ and $\Omega_{bc}$ spectra affects the masses of the lowest lying states by less than 20 MeV, but can be expected to have significant implications for other aspects of the phenomenology of these states, such as their semileptonic decay rates. A preliminary investigation of this has been the subject of a recent manuscript \cite{pragain}, in which it has been found that the mixing does indeed significantly affect some of the exclusive semileptonic decay channels. This mixing may also have important consequences for the electromagnetic and strong decays of these states.

\section{Conclusion and outlook}

We have applied a quark model of baryons to states containing one, two or three heavy quarks. There are a number of features of the model that distinguish it from other work of this type in the literature. All quarks contribute fully to the dynamics in the baryon, with no special approximations made for heavy quarks or heavy diquarks. The results of the model can therefore be used to examine how well such models approach the heavy-quark limit. Baryons with $J^P$ up to $\shlf^+$ are treated.

In the sector of states containing a single heavy quark, the results obtained are in good agreement with experimental observations. A number of experimental states without spin-parity assignments match several states in the model, but more data, on decays, as well as a theoretical treatment of such decays, are needed before model states can be paired with experimental states. Among the $\Xi_c$ and $\Xi_b$, the model predicts mixing angles between the sextet and antitriplet states that are small but not negligible for the lower lying states, but which tend to get larger for some excited states. The mixing between the two sets of states does not necessarily vanish as the mass of the heavy quark is increased. 

We have also examined these states to identify which of them belong in the spin-multiplets expected in HQET. Because we diagonalize a Hamiltonian matrix to give a number of states with the same spin and parity, mixing among states can make it difficult to identify the spin multiplets. Nevertheless, we were successful in identifying a number of multiplets, and noted that as the mass of the heavy quark increased, more multiplets could be identified, or came closer to meeting our `multiplet criterion'. In some cases, mixing between states meant that multiplets probably could never be identified, regardless of how large the mass of the heavy quark became. In the case of the $\Xi_c$ and $\Xi_b$, sextet-antitriplet mixing complicated the identification of the multiplets, allowing fewer to be isolated.

Among the baryons with three heavy quarks, our predictions have tended to be somewhat heavier than those of a number of authors, but not outrageously so. The main reason for this is the very small Coulomb interaction that results from our fits to the known baryons. Since none of the states containing three quarks have yet been seen, it is much too early to argue vigorously for one model over another. Among the $\Xi_{bc}$ and $\Omega_{bc}$, the model indicates that mixing between states comprising a scalar $bc$ diquark and those comprising an axial-vector $bc$ diquark is very strong, leading to very large mixing angles. In the language of the heavy quark expansion, $1/m_Q$ corrections will have large effects on the wave functions of the states in these sectors, even if the effects on the masses are small. The model further indicated that if such states are treated as being made of a $qc$ diquark, with $q=u,\,\,d$ or $s$, mixing between the scalar and axial-vector diquark could be more safely ignored, at least for the lowest lying states. For excited states, including those with negative parity, either choice of basis would be valid, but mixing was large. We have argued that the mixing in these states will have significant effects on their semileptonic decays, as well as on their electromagnetic and strong transitions. These investigations are left as possible future projects.

\acknowledgments

This work is supported by the Department of Energy, Office of Nuclear Physics,  under
contracts no. DE-AC02-06CH11357 (MP) and DE-AC05-06OR23177 (WR). WR is grateful to the Department of Physics, the College of Arts and Sciences and the Office of Research at Florida State University for partial support. The authors are grateful to J. Goity for useful discussions.

\appendix

\section{Wave Function Components\label{components}}

The spin-space components of the wave functions used are presented in Table \ref{wfcomponents}. The wave function components that are labelled as {\bf $\overline{3}$} or {\bf 1} are valid for any state whose flavor wave function is antisymmetric in the first two quarks, while those labelled as {\bf 6} or {\bf 3} are valid for any state whose flavor wave function is symmetric in the first two quarks. The notation in the table is
\beq
\left[\psi_{LM_Ln_{\rho}\ell_{\rho}n_{\lambda}\ell_\lambda}({\bf \rho}, {\bf 
\lambda})\chi_{S}(M-M_L)\right]_{J,M}\equiv \sum_{M_L}\< JM|LM_L, SM-M_L\>\psi_{LM_Ln_{\rho}
\ell_{\rho}n_{\lambda}\ell_\lambda}({\bf \rho}, {\bf 
\lambda})\chi_{S}(M-M_L),
\eeq
where $\chi_S$ is the spin wave function for a baryon with total quark spin $S$, and 
\beq
\psi_{LMn_{\rho}\ell_{\rho}n_{\lambda}\ell_\lambda}({\bf \rho}, {\bf 
\lambda}) = 
\sum_m\langle LM|\ell_{\rho}m,\ell_\lambda M-m\rangle\psi_{n_\rho \ell_\rho m}
({\bf \rho}) \psi_{n_\lambda \ell_\lambda M-m}({\bf \lambda}),
\eeq
is the spatial wave function.

\begin{center}
\begin{table}[h]
\caption{The spin-space wave function components that we use in our model. The notation is explained in the text.
\label{wfcomponents}}
\vspace{5mm}
\begin{tabular}{|c|c|c|c|}
\hline
$J^P$ & Component &\multicolumn{2}{c|}{Spin-Space Wave Function}\\\cline{3-4}
&& $f={\bf \overline{3}}$ or $f={\bf 1}$ & $f={\bf 6}$ or $f={\bf 3}$\\[+4pt]\hline\hline
\multirow{11}{*}{$\frac{1}{2}^+$} & $|1\rangle_f$ &$\psi_{000000}({\bf \rho}, {\bf \lambda}) \chi_{1/2}^\rho(M)$
&$\psi_{000000}({\bf \rho}, {\bf \lambda}) \chi_{1/2}^\lambda(M)$\\[+4pt]\cline{2-4}
&$|2\rangle_f$&$\psi_{001000}({\bf \rho}, {\bf \lambda})\chi_{1/2}^\rho(M)$ &$\psi_{001000}({\bf \rho}, {\bf \lambda})\chi_{1/2}^\lambda(M)$\\[+4pt]\cline{2-4}
&$|3\rangle_f$& $\psi_{000010}({\bf \rho}, {\bf \lambda})\chi_{1/2}^\rho(M)$
&$\psi_{000010}({\bf \rho}, {\bf \lambda})\chi_{1/2}^\lambda(M)$\\[+4pt]\cline{2-4}
&$|4\rangle_f$&$\psi_{000101}({\bf \rho}, {\bf \lambda})\chi_{1/2}^\lambda(M)$ &$\psi_{000101}({\bf \rho}, {\bf \lambda})\chi_{1/2}^\rho(M)$\\[+4pt]\cline{2-4}
&$|5\rangle_f$&$\left[\psi_{1M_L0101}({\bf \rho}, {\bf \lambda})
\chi_{3/2}^S(M-M_L)\right]_{1/2, M} $
&$\left[\psi_{1M_L0101}({\bf \rho}, {\bf \lambda})
\chi_{1/2}^\rho(M-M_L)\right]_{1/2, M}$\\[+4pt]\cline{2-4}
&$|6\rangle_f$&$\left[\psi_{1M_L0101}({\bf \rho}, {\bf \lambda})\chi_{3/2}^\lambda(M-M_L)\right]_{1/2, M}$ 
&$\left[\psi_{2M_L0200}({\bf \rho}, {\bf \lambda})\chi_{3/2}^S(M-M_L)\right]_{1/2, M}$\\[+4pt]\cline{2-4}
&$|7\rangle_f$& $\left[\psi_{2M_L0101}({\bf \rho}, {\bf \lambda})
\chi_{3/2}^S(M-M_L)\right]_{1/2, M}$&$\left[\psi_{2M_L0002}({\bf \rho}, {\bf \lambda})
\chi_{3/2}^S(M-M_L)\right]_{1/2, M}$\\[+4pt]\hline\hline
\multirow{13}{*}{$\frac{3}{2}^+$} & $|1\rangle_f$ &$\psi_{000101}({\bf \rho},{\bf \lambda})\chi_{3/2}^S(M)$
&$\psi_{000000}({\bf \rho}, {\bf \lambda}) \chi_{3/2}^S(M)$\\[+4pt]\cline{2-4}
&$|2\rangle_f$&$\left[\psi_{1M_L0101}({\bf \rho}, {\bf \lambda})\chi_{3/2}^S(M-M_L)\right]_{3/2,M}$ 
&$\psi_{001000}({\bf \rho}, {\bf \lambda})\chi_{3/2}^S(M)$\\[+4pt]\cline{2-4}
&$|3\rangle_f$& $\left[\psi_{1M_L0101}({\bf \rho}, {\bf \lambda})\chi_{1/2}^\lambda(M-M_L)\right]_{3/2,M}$
&$\psi_{000010}({\bf \rho}, {\bf \lambda})\chi_{3/2}^S(M)$\\[+4pt]\cline{2-4}
&$|4\rangle_f$&$\left[\psi_{2M_L0200}({\bf \rho}, {\bf \lambda})\chi_{1/2}^\rho(M-M_L)\right]_{3/2,M}$ 
&$\left[\psi_{1M_L0101}({\bf \rho}, {\bf \lambda})\chi_{1/2}^\rho(M-M_L)\right]_{3/2,M}$\\[+4pt]\cline{2-4}
&$|5\rangle_f$&$\left[\psi_{2M_L0101}({\bf \rho}, {\bf \lambda})
\chi_{3/2}^S(M-M_L)\right]_{3/2, M} $
&$\left[\psi_{2M_L0200}({\bf \rho}, {\bf \lambda})
\chi_{3/2}^S(M-M_L)\right]_{3/2, M}$\\[+4pt]\cline{2-4}
&$|6\rangle_f$&$\left[\psi_{2M_L0101}({\bf \rho}, {\bf \lambda})\chi_{1/2}^\lambda(M-M_L)\right]_{3/2, M}$ 
&$\left[\psi_{2M_L0200}({\bf \rho}, {\bf \lambda})\chi_{1/2}^\lambda(M-M_L)\right]_{3/2, M}$\\[+4pt]\cline{2-4}
&$|7\rangle_f$& $\left[\psi_{2M_L0002}({\bf \rho}, {\bf \lambda})
\chi_{1/2}^\rho(M-M_L)\right]_{3/2, M}$
&$\left[\psi_{2M_L0101}({\bf \rho}, {\bf \lambda})
\chi_{1/2}^\rho(M-M_L)\right]_{3/2, M}$\\[+4pt]\cline{2-4}
&$|8\rangle_f$&-&$\left[\psi_{2M_L0002}({\bf \rho}, {\bf \lambda})\chi_{3/2}^S(M-M_L)\right]_{3/2, M}$\\[+4pt]\cline{2-4}
&$|9\rangle_f$& - &$\left[\psi_{2M_L0002}({\bf \rho}, {\bf \lambda})
\chi_{3/2}^\lambda(M-M_L)\right]_{1/2, M}$\\[+4pt]\hline\hline
\multirow{8}{*}{$\frac{5}{2}^+$} & $|1\rangle_f$ &$\left[\psi_{1M_L0101}({\bf \rho}, {\bf \lambda}) 
\chi_{3/2}^S(M-M_L)\right]_{5/2,M}$
&$\left[\psi_{2M_L0101}({\bf \rho}, {\bf \lambda}) \chi_{1/2}^\rho(M-M_L)\right]_{5/2,M}$\\[+4pt]\cline{2-4}
&$|2\rangle_f$&$\left[\psi_{2M_L0101}({\bf \rho}, {\bf \lambda})\chi_{3/2}^S(M-M_L)\right]_{5/2,M}$ 
&$\left[\psi_{2M_L0200}({\bf \rho}, {\bf \lambda})\chi_{1/2}^S(M-M_L)\right]_{5/2,M}$\\[+4pt]\cline{2-4}
&$|3\rangle_f$& $\left[\psi_{2M_L0101}({\bf \rho}, {\bf \lambda})\chi_{1/2}^\lambda(M-M_L)\right]_{5/2,M}$
&$\left[\psi_{2M_L0200}({\bf \rho}, {\bf \lambda})\chi_{1/2}^\lambda(M-M_L)\right]_{5/2,M}$\\[+4pt]\cline{2-4}
&$|4\rangle_f$&$\left[\psi_{2M_L0200}({\bf \rho}, {\bf \lambda})\chi_{1/2}^\rho(M-M_L)\right]_{5/2,M}$ 
&$\left[\psi_{2M_L0002}({\bf \rho}, {\bf \lambda})\chi_{3/2}^S(M-M_L)\right]_{5/2,M}$\\[+4pt]\cline{2-4}
&$|5\rangle_f$&$\left[\psi_{2M_L0002}({\bf \rho}, {\bf \lambda})
\chi_{1/2}^\rho(M-M_L)\right]_{1/2, M}$
&$\left[\psi_{2M_L0002}({\bf \rho}, {\bf \lambda})
\chi_{1/2}^\lambda(M-M_L)\right]_{5/2, M}$\\[+4pt]\hline\hline
\multirow{3}{*}{$\frac{7}{2}^+$} & $|1\rangle_f$ &$\left[\psi_{2M_L0101}({\bf \rho}, {\bf \lambda})\chi_{3/2}^S(M-M_L)\right]_{7/2,M}$
&$\left[\psi_{2M_L0200}({\bf \rho}, {\bf \lambda}) \chi_{3/2}^S(M-M_L)\right]_{7/2,M}$\\[+4pt]\cline{2-4}
&$|2\rangle_f$& - &$\left[\psi_{2M_L0002}({\bf \rho}, {\bf \lambda})\chi_{3/2}^S(M-M_L)\right]_{7/2,M}$\\[+4pt]\hline\hline
\multirow{4}{*}{$\frac{1}{2}^-$} & $|1\rangle_f$ &$\left[\psi_{1M_L0100}({\bf \rho}, {\bf \lambda}) \chi_{3/2}^S(M-M_L)\right]_{1/2,M}$
&$\left[\psi_{1M_L0100}({\bf \rho}, {\bf \lambda}) \chi_{1/2}^\rho(M-M_L)\right]_{1/2,M}$\\[+4pt]\cline{2-4}
& $|2\rangle_f$ &$\left[\psi_{1M_L0100}({\bf \rho}, {\bf \lambda}) \chi_{3/2}^\lambda(M-M_L)\right]_{1/2,M}$
&$\left[\psi_{1M_L0001}({\bf \rho}, {\bf \lambda}) \chi_{3/2}^S(M-M_L)\right]_{1/2,M}$\\[+4pt]\cline{2-4}
& $|3\rangle_f$ &$\left[\psi_{1M_L0001}({\bf \rho}, {\bf \lambda}) \chi_{1/2}^\rho(M-M_L)\right]_{1/2,M}$
&$\left[\psi_{1M_L0001}({\bf \rho}, {\bf \lambda}) \chi_{1/2}^\lambda(M-M_L)\right]_{1/2,M}$\\[+4pt]\hline\hline
\multirow{4}{*}{$\frac{3}{2}^-$} & $|1\rangle_f$ &$\left[\psi_{1M_L0100}({\bf \rho}, {\bf \lambda}) \chi_{3/2}^S(M-M_L)\right]_{3/2,M}$
&$\left[\psi_{1M_L0100}({\bf \rho}, {\bf \lambda}) \chi_{1/2}^\rho(M-M_L)\right]_{3/2,M}$\\[+4pt]\cline{2-4}
& $|2\rangle_f$ &$\left[\psi_{1M_L0100}({\bf \rho}, {\bf \lambda}) \chi_{3/2}^\lambda(M-M_L)\right]_{3/2,M}$
&$\left[\psi_{1M_L0001}({\bf \rho}, {\bf \lambda}) \chi_{3/2}^S(M-M_L)\right]_{3/2,M}$\\[+4pt]\cline{2-4}
& $|3\rangle_f$ &$\left[\psi_{1M_L0001}({\bf \rho}, {\bf \lambda}) \chi_{1/2}^\rho(M-M_L)\right]_{3/2,M}$
&$\left[\psi_{1M_L0001}({\bf \rho}, {\bf \lambda}) \chi_{1/2}^\lambda(M-M_L)\right]_{3/2,M}$\\[+4pt]\hline\hline
$\frac{5}{2}^-$ & $|1\rangle_f$ &$\left[\psi_{1M_L0100}({\bf \rho}, {\bf \lambda}) \chi_{3/2}^S(M-M_L)\right]_{5/2,M}$
&$\left[\psi_{1M_L0001}({\bf \rho}, {\bf \lambda}) \chi_{3/2}^S(M-M_L)\right]_{5/2,M}$\\\hline
\end{tabular}
\end{table}
\end{center}

\section{HQET Multiplets in Terms of Quark Model Quantum Numbers \label{transstates}}

In this appendix, we give the explicit structures for the HQET multiplets expected, in terms of the quark model quantum numbers that we use. For the HQET states, the notation is $\left|j, J\right>$, where $j$ is the total angular momentum of the light component of the baryon, and $J$ is the total angular momentum of the baryon. For the quark model states, the notation is $\left|L,s_{12},S,J\right>$, where $L$ is the total orbital angular momentum in the baryon ${\bf L}={\bf \ell}_\rho+{\bf \ell}_\lambda$, $s_{12}$ is the spin of the light pair of quarks (we treat the heavy quark as being the third quark in the baryon), and $S$ is the total spin of all the quarks in the baryon. For each degenerate pair of heavy baryon states, there are several sets of quark model quantum numbers that lead to the same state. In addition, any state shown may also refer to radially excited states. We list all the ones relevant to the present model. We note that the expressions we write are valid for baryons of either parity.

We begin with singlet states. There are two ways in which these can be constructed, and these are
\beqy \label{spinsinglet}
\left|0,\hlf\right>&=&\left|0,0,\hlf,\hlf\right>,\nonumber\\
\left|0,\hlf\right>&=&\sqrt{\frac{2}{3}}\left|1,1,\thlf,\hlf\right>-
\sqrt{\frac{1}{3}}\left|1,1,\hlf,\hlf\right>.
\eeqy

There are four ways to construct the states of the $(\hlf,\,\,\thlf)$ multiplet of either parity. The states are
\beqy
\left|1,\hlf\right>&=&\left|0,1,\hlf,\hlf\right>,\,\,\,\,\left|1,\thlf\right>=\left|0,1,\thlf,\thlf\right>,\nonumber\\
\left|1,\hlf\right>&=&\left|1,0,\hlf,\hlf\right>,\,\,\,\,\left|1,\thlf\right>=\left|1,0,\hlf,\thlf\right>,\nonumber\\
\left|1,\hlf\right>&=&\sqrt{\frac{2}{3}}
\left|1,1,\hlf,\hlf\right>+\frac{1}{\sqrt{3}}\left|1,1,\thlf,\hlf\right>,\,\,\,\,
\left|1,\thlf\right>=-\frac{1}{\sqrt{6}}
\left|1,1,\hlf,\thlf\right>+\sqrt{\frac{5}{6}}\left|1,1,\thlf,\thlf\right>,\nonumber\\
\left|1,\hlf\right>&=&\left|2,1,\thlf,\hlf\right>,\,\,\,\,
\left|1,\thlf\right>=\frac{1}{\sqrt{2}}\left(
-\left|2,1,\hlf,\thlf\right>+\left|2,1,\thlf,\thlf\right>\right),\nonumber\\
\eeqy

There are three ways to construct the states of the $(\thlf,\,\,\fhlf)$ multiplet of either parity. Note that here, we limit the possibilities to states with $L\le2$. The states are
\beqy
\left|2,\thlf\right>&=&\left|2,0,\hlf,\thlf\right>,\,\,\,\,
\left|2,\fhlf\right>=\left|2,0,\hlf,\fhlf\right>,\nonumber\\
\left|2,\thlf\right>&=&\sqrt{\frac{5}{6}}
\left|1,1,\hlf,\thlf\right>+\frac{1}{\sqrt{6}}\left|1,1,\thlf,\thlf\right>,\,\,\,\,
\left|2,\fhlf\right>=\left|1,1,\thlf,\fhlf\right>,\nonumber\\
\left|2,\thlf\right>&=&\frac{1}{\sqrt{2}}\left(
\left|2,1,\hlf,\thlf\right>+\left|2,1,\thlf,\thlf\right>\right),\,\,\,\,
\left|2,\fhlf\right>=\frac{\sqrt{7}}{3}
\left|2,1,\hlf,\fhlf\right>-\frac{\sqrt{2}}{3}\left|2,1,\hlf,\fhlf\right>.
\eeqy

Limiting the value of $L$ to less than two, there is only one way to construct the states of the $(\fhlf,\,\,\shlf)$ multiplet, and that is
\beqy
\left|3,\fhlf\right>&=&\frac{\sqrt{2}}{3}
\left|2,1,\hlf,\fhlf\right>+\frac{\sqrt{7}}{3}\left|2,1,\thlf,\fhlf\right>,\,\,\,\,
\left|3,\shlf\right>=\left|2,1,\thlf,\shlf\right>
\eeqy

%%%%%%%%%%%%%%%%%%%%%%%%%%%%%%%%%%%%%%%%%%%%%%%%%%%%%%%%%%%%%%%%%%%
%%%%%%%%%%%%%%%%%%%%%%%%%%%%%%%%%%%%%%%%%%%%%%%%%%%%%%%%%%%%%%%%%%%
%%%%%%%%%%%%%%%%  MACRO FOR THE REFERENCES  %%%%%%%%%%%%%%%%%%%%%%%
%%%%%%%%%%%%%%%%%%%%%%%%%%%%%%%%%%%%%%%%%%%%%%%%%%%%%%%%%%%%%%%%%%%
%%%%%%%%%%%%%%%%%%%%%%%%%%%%%%%%%%%%%%%%%%%%%%%%%%%%%%%%%%%%%%%%%%%
\newif\ifmultiplepapers
\def\beginpapers{\multiplepaperstrue}
\def\endpapers{\multiplepapersfalse}  
\def\journal#1&#2(#3)#4{\rm #1~{\bf #2}\unskip, \rm  #4 (19#3)}
\def\trjrnl#1&#2(#3)#4{\rm #1~{\bf #2}\unskip, \rm #4 (19#3)}
\def\baps{\journal {Bull.} {Am.} {Phys.} {Soc.}&}
\def\jap{\journal J. {Appl.} {Phys.}&}
\def\prl{\journal {Phys.} {Rev.} {Lett.}&}
\def\pl{\journal {Phys.} {Lett.}&}
\def\pr{\journal {Phys.} {Rev.}&}
\def\np{\journal {Nucl.} {Phys.}&}
\def\rmp{\journal {Rev.} {Mod.} {Phys.}&}
\def\jmp{\journal J. {Math.} {Phys.}&}
\def\rmm{\journal {Revs.} {Mod.} {Math.}&}
\def\jetp{\journal {J.} {Exp.} {Theor.} {Phys.}&}
\def\sjetp{\trjrnl {Sov.} {Phys.} {JETP}&}
\def\dokl{\journal {Dokl.} {Akad.} Nauk USSR&}
\def\spd{\trjrnl {Sov.} {Phys.} {Dokl.}&}
\def\tmf{\journal {Theor.} {Mat.} {Fiz.}&}
\def\snp{\trjrnl {Sov.} J. {Nucl.} {Phys.}&}
\def\hpa{\journal {Helv.} {Phys.} Acta&}
\def\yf{\journal {Yad.} {Fiz.}&}
\def\zp{\journal Z. {Phys.}&}
\def\anp{\journal {Adv.} {Nucl.} {Phys.}&}
\def\ap{\journal {Ann.} {Phys.}&}
\def\am{\journal {Ann.} {Math.}&}
\def\nc{\journal {Nuo.} {Cim.}&}
\def\etal{{\sl et al.}}
\def\pre{\journal {Phys.} {Rep.}&}
\def\pca{\journal Physica (Utrecht)&}
\def\prs{\journal {Proc.} R. {Soc.} London &}
\def\jcp{\journal J. {Comp.} {Phys.}&}
\def\pna{\journal {Proc.} {Nat.} {Acad.}&}
\def\jpg{\journal J. {Phys.} G (Nuclear Physics)&}
\def\fort{\journal {Fortsch.} {Phys.}&}
\def\jfa{\journal {J.} {Func.} {Anal.}&}
\def\cmp{\journal {Comm.} {Math.} {Phys.}&}
%%%%%%%%%%%%%%%%%%%%%%%%%%%%%%%%%%%%%%%%%%%%%%%%%%%%%%%%%%%%%%%%%%%

\end{document}